\documentclass[aps,prd,reprint]{revtex4-1}
\usepackage{graphicx}
\usepackage{amsfonts}
\usepackage{amssymb}
\usepackage{amsmath}
\usepackage{longtable}
\newcommand\gtsim{\mathrel{\lower0.6ex\hbox{$\buildrel {\textstyle >}\over {\scriptstyle \sim}$}}}
\newcommand\ltsim{\mathrel{\lower0.6ex\hbox{$\buildrel {\textstyle <}\over {\scriptstyle \sim}$}}}
\newcommand\HST{{\it HST }}
\newcommand{\farcs}{\mbox{\ensuremath{.\mkern-4mu^\prime}}}

\usepackage{enumerate}
\begin{document}
\bibliographystyle{apsrev}

\title{Limits on Spacetime Foam}
\author{Wayne A. Christiansen}
\affiliation{Department of Physics and Astronomy, University of North Carolina,
Chapel Hill, NC 27599}
\author{David J. E. Floyd}
\affiliation{AAO/OCIW Magellan Fellow. Current address: 
School of Physics, University of Melbourne, Victoria 3010, Australia}
\author{Y. Jack Ng}
\affiliation{Department of Physics and Astronomy, University of North Carolina,
Chapel Hill, NC 27599}
\author{Eric S. Perlman}
\affiliation{Physics and Space Sciences Department, 
Florida Institute of Technology, Melbourne, FL 32901}

\begin{abstract}
Plausibly spacetime is ``foamy'' on small distance scales, due to quantum
fluctuations. We elaborate on the proposal to detect spacetime foam by
looking for seeing disks in the images of distant quasars and AGNs.
This is a null test in the sense that the continued presence of unresolved
``point'' sources at the milli-arc second level in samples of distant
compact sources puts severe constraints on theories of quantized spacetime
foam at the Planckian level. We discuss the geometry of foamy spacetime, and 
the appropriate distance measure for calculating the expected angular broadening.
We then deal with recent data and the constraints they put on spacetime foam models. 
While time lags from distant pulsed sources such as GRBs have been posited as
a possible test of spacetime foam models, we demonstrate that the time-lag effect is 
rather smaller than has been calculated, due to the equal probability of positive and 
negative fluctuations in the speed of light inherent in such models. 
Thus far, images of high-redshift quasars from the Hubble Ultra-Deep Field (UDF)
provide the most stringent test of spacetime foam theories. While random walk 
models ($\alpha = 1/2$) have already been ruled out, 
the holographic ($\alpha=2/3$) model remains viable. 
Here $\alpha \sim 1$ parametrizes the different spacetime foam models 
according to which the fluctuation of a distance $l$ is given by
$\sim l^{1 - \alpha} l_P^{\alpha}$ with $l_P$ being the Planck length.
Indeed, we see a slight wavelength-dependent blurring in the UDF 
images selected for this study. Using existing data in the 
{\it Hubble Space Telescope (HST)} archive we find it is impossible to rule out the 
$\alpha=2/3$ model, but exclude all models with $\alpha<0.65$.
By comparison, current GRB time lag observations only exclude models with $\alpha<0.3$.
\end{abstract}


\maketitle

%I. {\bf Introduction}
\section{Introduction}
\label{sec-intro}
Even at the minute scales of distance and duration examined with
increasingly discriminating instruments, spacetime still appears
to be smooth and structureless. 
But, like everything else, (plausibly) spacetime is subject to quantum
fluctuations. So we expect that spacetime, probed at a small enough scale,
will appear complicated --- something akin in complexity to a turbulent
froth that~\cite{whe63}
has dubbed ``quantum foam,'' also known as ``spacetime
foam.'' But how large are the fluctuations in the fabric of
spacetime? %\cite{ng03b} 
How small a scale is small enough so that the foaminess of spacetime 
manifests itself? 
To quantify the problem, let us recall that, if spacetime indeed
undergoes quantum fluctuations, there will be an intrinsic limitation to
the accuracy with which one can measure a macroscopic distance, for that distance
fluctuates. Denoting the fluctuation of a distance $l$ by $\delta l$, on
general grounds, we expect $\delta l \gtrsim N l^{1 - \alpha} l_P^{\alpha}$,
\cite{ng03b} where $N$ is a numerical factor $\sim 1$ and 
$l_P = \sqrt{\hbar G/c^3}$ is the Planck length,
the characteristic length scale in quantum gravity, and we have denoted the
Planck constant, gravitational constant and the speed of light by $ \hbar$,
$G$ and $c$ respectively. 
It is important to note that $\delta l$ cannot be defined without reference to the
macroscopic distance, $l$ (i.e., $\delta l$ is not defined locally).
The parameter $\alpha
\lesssim 1$ specifies the different spacetime foam models.
We note that smaller values of $\alpha$ necessarily lead to larger distance 
fluctuations and, hence, such models are easier to test observationally.
%In fact a major goal of this paper is to set an observational 
%lower limit on the $\alpha-$parameter.

Because the Planck length ($\sim 10^{-33}$ cm) is so incredibly small, we
need an astronomically large distance $l$ for its fluctuation $\delta l$ to be
detectable.
Even so, measurement of $\delta l$ is not trivial. It must be calibrated against
a known length standard. The most obvious calibrator is the wavelength 
$\lambda$, of the light received from a distant source. Thus, in principle, 
distance fluctuations $\pm \delta l$, imply phase fluctuations
$\pm \Delta \phi = \pm 2 \pi \delta l / \lambda$ (c.f.~\cite{lie03,rag03,ng03a}).
More recently, Christiansen et al.~\cite{chr06}, henceforth CNvD, suggested
that fluctuations in the direction of the local wave-vector, 
$\pm \delta \psi \equiv
\pm \Delta \phi /(2 \pi) = \pm \delta l / \lambda$ could possibly be detected
as ``halos'' in images of distant ``point sources''. However, it must be noted 
that using the wavelength of received light to bootstrap a calibration for
measuring $\delta l$, necessarily involves interferometry and/or wave optics
such as Strehl analysis (see section~\ref{sec-meas}).
%\cite{lie03,rag03,ng03a} Recently
%it was suggested by Christiansen et al~\cite{chr06}, henceforth CNvD,
%that spacetime foam might be uncovered by looking for unresolved cores
%in the images of distant quasars. We start with a detailed
%summary of this proposal and expound on some of its potential problems. 
The point is that, due to
quantum foam-induced fluctuations in the phase velocity of an incoming light
wave from a distant point source, the wave front itself develops a small
scale ``cloud of uncertainty'' equivalent to a ``foamy'' structure. 
%because parts of the wave-front lag while other
%parts advance . 
This results in the wave vector, upon detection, acquiring a jitter in
direction with an angular spread of the order of
$\delta \psi$. 
%where $\Delta \phi = 2 \pi \delta l / \lambda
%= 2 \pi N l^{1-\alpha} l_P^{\alpha} / \lambda$
%is the fluctuation in the phase of the electromagnetic wave with wavelength
%$\lambda$ after traveling a distance $l$ from the distant source.\cite{chr06} 
%We note that both the phase and the angular 
%fluctuations are derived from the
%fundamental uncertainty in distance, 
%$\delta l$. The latter, as shown by
%CNvD, is based on the {\it conjecture} that 
%$\delta l$ is isotropic at the pointof observation.
In effect, spacetime foam creates a ``seeing disk'' whose angular diameter is
%\cite{chr06}
\begin{equation}
\delta \psi
= N (\frac {l}{\lambda})^{1 - \alpha} (\frac {l_P}{\lambda})^{\alpha}. 
\label{eq0}
\end{equation}
For a telescope or
interferometer with baseline length $D$, this
means that dispersion (on the order of $\delta \psi$ in the normal
to the wave front) will be recorded as a spread in the angular size of a
distant point source, causing a reduction in the Strehl ratio, and/or the
fringe visibility when
$\delta \psi \sim \lambda / D$ for a diffraction limited telescope.
Thus, in principle, for arbitrarily large distances spacetime foam sets a lower 
limit on the observable angular size of a source at a given wavelength 
$\lambda$. Furthermore, the disappearance
of ``point sources'' will be strongly wavelength dependent happening first at short
wavelengths. 
%In that respect X-ray imaging by {\it Chandra} may be definitive if some
%technical problems can be solved as discussed below.
In fact, as we show below, the frequency dependence of blurring currently offers the 
best chance to detect the effects of spacetime foam. 

As discussed by CNvD,
there are three major theoretical models for specifying the cumulative effects of
spacetime foam:
\begin{itemize}
\item The random-walk model~\cite{ame99} 
in which the spacetime foam effects grow like a one-dimensional random walk, 
corresponding to $\alpha = 1/2, N = 1$. 
\item The holographic model~\cite{kar66,ng94,ng95} which is consistent
with the holographic principle~\cite{tho93,sus95} where the
information content in any three-dimensional
region of space can be encoded on a two-dimensional surface surrounding the
region of interest, corresponding to $\alpha = 2/3$ and $0.97 \leq N \leq 1.9$
as discussed in section~\ref{sec-dfe}.
\item An anti-correlation model~\cite{mis73}
in which there are no cumulative effects, so the
distance fluctuation remains simply the Planck length, corresponding to 
$\alpha = 1, N = 1$.
\end{itemize}

The random walk model can be ruled out on the basis of \HST observations 
of relatively nearby quasars (distance $\sim$ 1 Gpc) according to CNvD. The
predicted spacetime foam induced seeing disks ($\sim$ few $\times 10^{-4}$ 
radians) clearly exceed the \HST resolution even in the IR waveband 
at 800 nm ($3.3 \times 10^{-7}$ radians). 
Although the holographic model's seeing disk ($\sim 10^{-8}$ radians) is still 
an order of magnitude smaller than the nominal IR \HST resolution, at larger
distances (redshifts $z \gtrsim 1$) and shorter wavelengths \HST may have 
sufficient resolution to test this model also.
On the other hand the anti-correlation model with 
$\delta \psi \sim l_P/ \lambda$ will likely remain untestable by 
astronomical means since the effect is vanishingly small and independent of 
distance.

Testing models of spacetime foam 
requires large distances to maximize the effect which, therefore, 
requires that cosmological effects must be taken into account.
The fundamental cosmological distance measure is the {\em co-moving distance}
(c.f.~\cite{hog00}). However, other distance measures may be used, depending
on circumstances. For example, the 
{\em luminosity distance} is commonly used to interpret observations of high 
redshift objects (as used, for example, by
\cite{ste07}, in his study of spacetime foam angular broadening).
However luminosity distance is inappropriate for studying spacetime foam
induced angular broadening. 
The reason is that the luminosity distance has a built-in
energy dependence because it is based on the flux detected by a distant
observer. Therefore, the redshift dependence of the luminosity distance
explicitly includes a correction for the diminished energy received from
distant sources. That enhances the expected angular broadening 
(c.f. Eq.~(\ref{eq0})).
However, the angular broadening caused by spacetime foam
does not explicitly depend on flux. 
Another distance measure, the angular diameter distance, may be appropriate for
observations involving lensing or scattering. 
However, spacetime foam angular uncertainties are not associated with scattering
or lensing (see section~\ref{sec-meas} for detailed discussion). 
Quantum foam is, by definition, a property of the fabric of spacetime itself, 
and not an extrinsic property measured against the background geometry. 
Therefore the appropriate cosmological distance measure is the 
fundamental {\em co-moving distance} which we shall use in our subsequent calculations.

In addition, other factors also make the search for observable effects more
complicated for the following reasons:
\begin{enumerate}
\item Sensitivity: More distant sources are fainter. This means that, as far
as interferometry is concerned, the lack of observed fringes may simply be
due to the lack of sufficient flux rather than the possibility that the
instrument has resolved a spacetime foam generated halo. This is probably
the most serious limitation on experiments to search for spacetime
foam. As discussed above, the effect is cumulative, requiring large
distances to appear. But large distances inevitably mean faint sources.
\item Atmospheric turbulence:
%Sensitivity There is also the parallel problem of how to distinguish the small effects
%due to quantum foam from similar effects due to the turbulence of the
%Earth's atmosphere. 
Integration times must be shorter than atmospheric fluctuation times. 
Thus again what one needs is to find a bright but sufficiently distant
external source, for then the integration times may be short enough to allow
one to remove or at least minimize the atmospheric effects.
Of course, the latter problem does not intrude for space-based observatories
such as \HST or {\it Chandra}, nor for an interferometer.
\item Masking: This may originate under two distinct circumstances:
\begin{enumerate}[(a)]
\item The source may have an intrinsic structure comparable in angular size to
the expected spacetime foam halo; 
\item Any competing scattering mechanism in the intervening medium between 
source and observer (e.g., gravitational wave scattering of light, 
micro-lensing by machos or dark matter, etc.)
\end{enumerate}
\end{enumerate}
Concerning (a), distant sources may have extended structures. Thus if
fringes are not seen,
it may be an indication that the source quasar has an intrinsic core
structure whose angular size is comparable to, or larger than, the spacetime
foam halo. Since we do not know the physical scales of all components of
AGN and quasars we may find that spacetime foam effects are masked by
intrinsic structures. However, since the angular size of intrinsic structures 
decreases like $l^{-1}$ (in a flat universe) whereas spacetime foam 
halos {\em increase} as $l^{1 - \alpha}$, statistical samples can, in principle, 
separate the two effects.

The bottom line is that a number of modern telescopes including the 
Very Large Telescope {\it VLTI}, \HST and {\it Chandra}
are on the verge of testing theories of spacetime foam
(see, e.g.,~\cite{ste07}), providing a sufficiently large sample of
sufficiently bright compact sources can be developed. 
This is where future large aperture telescopes will prove to be crucial.
The test is simply a question of
the detection of a reduction in Strehl ratio or correspondingly a reduction
in interferometer fringe amplitude.
It is not a question of mapping the structure of the
predicted seeing disk. Knowing the real sensitivity
limits of the telescopes involved
will enable us to search the catalogs to see if we can find some quasars
that exceed the sensitivity limits.

One of the purposes of this paper is to give a more quantitative analysis of
the proposal just described. For that, the order of magnitude estimate for
$\delta l$ available so far is not enough. In the next section, we will
calculate the numerical factor in the $\delta l$ expression. 

A recent observation~\cite{abdo09} of a noticeable spread in arrival times 
for high energy gamma rays from distant gamma ray bursts has prompted 
the suggestion that it is unlikely that fuzziness of radio or optical images of 
distant extragalactic sources would be observed. 
The latter observation is exactly what this paper is about, thus we need 
to show that we disagree with the suggestion. Our detailed argument 
(based on previous work~\cite{ng08}) is given in section~\ref{sec-grb}.

In section~\ref{sec-meas}, we present a more detailed discussion of wave 
vector uncertainty in spacetime foam models. We discuss specifics of the 
necessary null tests for spacetime foam models as well as the reasons 
why spacetime foam halos cannot be interpreted as resulting from multiple 
scattering events.

So far, in our attempts to detect spacetime foam, only the \HST images of a few
quasars/AGNs~\cite{per02,rag03}
have been analyzed~\cite{lie03,ng03a,chr06,ste07}. In section~\ref{sec-obstest}, 
we show the size of expected spacetime foam broadening for various 
telescopes in a variety of wavebands. We then concentrate on the 
optical regime. Noting that the {\it VLTI} will
not be in full operation for some time, we 
concentrate our attention in this paper on an analysis of quasar images in the \HST archive.
This is done in section~\ref{sec-HST}. 
We discuss results and conclusions in section~\ref{sec-conc}.\\

%II. {\bf Distance Fluctuation Expression}
\section{Distance Fluctuation Expression}
\label{sec-dfe}
In this section, we will calculate the numerical factor $N$ in the distance
fluctuation expression for 
$\delta l \gtrsim N l_P^{\alpha} l^{1 - \alpha}$. 
\begin{enumerate}[(a)]
\item For the random walk model with
step length $l_P$, taking $M$ steps with a probability $p$~~($q$) of taking a forward
(backward) step, the mean square displacement is given by 
$(\delta l)^2 = 4 M p q l_P^2$. For the distance $l$, the number of steps is
$M = l/l_P$. Therefore, with an unbiased random walk, $p = q =1/2$, we get
an root-mean-square displacement of $\delta l = l^{1/2} l_P^{1/2}$, so $N = 1$.
\item Obviously for the anti-correlation model, since $\delta l$ is always equal to 
$l_P$, again $N = 1$. 
\item For the holographic model, we now have an explicit calculation which sets 
narrow limits for $N$. We use four different methods and their consistency 
with each other (to within a factor of two or so, as
we will show) bodes well for the robustness of our result. The ingredients
behind all the approaches are black hole physics and quantum mechanics.
\end{enumerate}

%1. Quantum computation
\subsection{Quantum computation}
This method~\cite{llo04,gio04}
relies on the fact that quantum fluctuations of spacetime manifest
themselves in the form of uncertainties in the geometry of
spacetime. Hence the structure of spacetime foam can be inferred from the
accuracy with which we can measure that geometry. Let us consider mapping
out the geometry of spacetime for a spherical volume of radius $l$ over the
amount of time $T = 2l/c$ it takes light to cross the
volume. One way to do this is to fill the space with clocks, exchanging
signals with other clocks and measuring the signals' times of arrival. This
process of mapping the geometry of spacetime is a kind of computation, in
which distances are gauged by transmitting and processing information.
The total number of operations, including the ticks of the clocks and
the measurements of signals, is bounded by the Margolus-Levitin
theorem~\cite{mar98}
in quantum computation, which stipulates that the rate of operations for any
computer cannot exceed the amount of energy $E$ that is available for
computation
divided by $\pi \hbar/2$. A total mass $M$ of clocks then
yields, via the Margolus-Levitin theorem, the bound on the total number of
operations given by $(2 M c^2 / \pi \hbar) \times 2 l/c$. But to prevent
black hole formation, $M$ must be less than $l c^2 /2 G$. Together, these
two limits imply that the total number of operations that can occur in a
spatial volume of radius $l$ for a time period $2 l/c$ is no greater than
$2 (l/l_P)^2 / \pi$. To maximize spatial resolution, each clock must tick
only once during the entire time period. If we regard the operations
partitioning the spacetime volume into ``cells'', then on the average each
cell
occupies a spatial volume no less than $(4 \pi l^3 / 3) / (2 l^2 /\pi l_P^2)
= 2 \pi^2 l l_P^2 /3$, yielding an average separation between neighboring
cells
no less than $(2 \pi^2 /3)^{1/3} l^{1/3} l_P^{2/3}$. This spatial
separation is interpreted as the average minimum uncertainty in the
measurement of a distance $l$, that is, $\delta l \gtrsim 1.9 l^{1/3}
l_P^{2/3}$; thus, $N = 1.9$.\\

%2. Heisenberg's Uncertainty Principle
\subsection{Heisenberg's Uncertainty Principle}
This method involves an estimate of the maximum number of (massless)
particles that can be put inside a spherical region of radius $l$.
Heisenberg's
uncertainty principle can be invoked to give the minimum momentum of each
particle to be ${1 \over 2} \times (\hbar / 2 l)$, or in other words, the
minimum energy of each particle is no less than ${1 \over 4} \times \hbar
c/l$.
To prevent black hole formation, the total energy is bounded by ${1 \over 2}
\times l c^4 /G$. Thus the total number of particles is no greater than $2
(l/l_P)^2$. This yields $ \delta l \gtrsim ({2 \pi \over 3})^{1/3} \times
l^{1/3} l_P^{2/3} = 1.3 l^{1/3} l_P^{2/3}$; thus, $N = 1.3$.\\

%3. Holographic Principle
\subsection{Holographic Principle}
This method~\cite{ng00,ng01}
is based on the upper bound on the number $I$
of degrees of freedom that can be put in a spherical volume of radius $l$.
According to the holographic principle,\cite{tho93,sus95}
the entropy $S$ in the sphere is
bounded by $S/ k_B \lesssim {1 \over 4} 4 \pi l^2 / l_P^2$, where $k_B$ is
the Boltzman constant. Recalling that $e^{S/k_B} = 2^{I}$, we get
$I \lesssim \pi (l/l_P)^2 /ln 2$. The average separation between
neighboring
degrees of freedom then yields $\delta l \gtrsim ({4 ln 2 \over 3})^{1/3}
l^{1/3} l_P^{2/3} = 0.97 l^{1/3} l_P^{2/3}$; thus, $N = 0.97$.\\

%4. The Wigner-Salecker Gedanken Experiment
\subsection{The Wigner-Salecker Gedanken Experiment}
In this method~\cite{wig57,sal58,ng94,ng95}
we conduct an ideal thought experiment to measure a distance $l$
between a clock and a mirror, By sending a light signal from the clock to
the
mirror in a timing experiment, we can determine $l$. The clock's and the
mirror's positions jiggle according to Heisenberg's uncertainty principle,
resulting in an uncertainty $\delta l$ in the measurement of $l$. 

Let us
concentrate on the fluctuation of the distance $l$ due to the jiggling of
the clock's position. If the clock of mass $m$ has a linear spread $\delta
l$ (and hence an uncertainty of speed, via Heisenberg's uncertainty
principe, given by $\hbar /( 2 m \delta l)$) when the light
signal leaves the clock, then its position spread grows to $\delta l + \hbar
{1 \over 2m \delta l} {2l \over c}$ when the light signal
returns to the clock, with the minimum at $\delta l = \hbar l / mc$. Thus
quantum mechanics alone would suggest using a massive clock. Now consider
the clock to be light-clock consisting of a spherical cavity of diameter
$d$,
surrounded by a mirror wall between which bounces a beam of light. For the
uncertainty in distance not to exceed $\delta l$, the clock must tick off
time
fast enough so that $d/c \lesssim \delta l /c$. But $d$ must be larger than
twice the Schwarzschild radius $2Gm/c^2$. These two requirements imply
$\delta l \gtrsim 4Gm/c^2$ which combines with the requirement from quantum
mechanics to yield $(\delta l)^3 \gtrsim 4 l l_P^2$ (independent of the
mass $m$ of the clock), or $\delta l \gtrsim 1.6 l^{1/3} l_P^{2/3}$; thus, 
$N = 1.6$.
A remark is in order: taking into account also the jiggling of the mirror's
position can only enhance $\delta l$.  As another remark, we note that the 
problem of distance measurement has been discussed in \cite{ellis92} in
non-critical string theory.

Thus, all four different methods yield very similar results. (Part of the
relatively small discrepancy can be traced to the mismatch between the
geometries of spheres and
cubes.) This is not surprising as our result depends not on any particular
theory of quantum gravity, but only on the quantum nature of measurements.
%The ``average'' fluctuation of a distance $l$ is given by 
%$\delta l \gtrsim 1.5 l^{1/3} l_P^{2/3}$.\\
In the discussion below, we will use the preceding arguments to set the 
allowable range for $N$ in the holographic model as : $0.97 \leq N \leq 1.9$.\\

%III. {\bf High Energy Gamma Rays from Distant GRB} 
\section{Spread in Arrival Times from Distant Pulsed Sources}
\label{sec-grb}
In this section we will discuss the technique of using the spread in arrival times of
photons as a possible technique for detecting spacetime foam. We are motivated by the
interesting detection of a minimal spread in the arrival times of high energy
photons from distant GRB reported by~\cite{abdo09}. 
While useful in putting a limit on the variation of
the speed of light of a definite sign, this technique is far less useful
than the measured angular size in constraining the
degree of fuzziness of spacetime in the spacetime
foam models that we consider in this paper. The
reason is that spacetime foam models predict that
the speed of light fluctuates (i.e., the fluctuation
takes on $\pm$ sign with equal probability): at one
instant a particular photon is faster than the average
of the other photons, but at the next instant it is
slower. We will show that the cumulative effect of
the fluctuations is too small to be detectable~\cite{ng08},
even for high-energy gamma rays that have
travelled $\sim 7$ billion light-years halfway across the
Universe~\cite{abdo09}, for most of the
spacetime foam models.

First we need to examine the cumulative effects~\cite{ng03a}
of spacetime fluctuations over a large distance. Consider a distance $l$,
and divide it into $l/ \lambda$ equal
parts each of which has length $\lambda$ ($\lambda$ can be as small as
$l_P$).
If we start with $\delta \lambda$ from each part, the question is how do
the $l/ \lambda$ parts
add up to $\delta l$ for the whole distance $l$. In other words, we want
to find
the cumulative factor $\mathcal{C}$ defined by
$\delta l = \mathcal{C}\, \delta \lambda$,
Since $\delta l \sim l^{1 - \alpha} l_P^{\alpha}$ and
$\delta \lambda \sim {\lambda}^{1 - \alpha} l_P^{\alpha}$, the result is
$\mathcal{C} = \left(\frac{l}{\lambda}\right)^{1 - \alpha}$. Note that
the cumulative factor is {\em not} linear in $(l/\lambda)$, i.e.,
$\frac{\delta l}{\delta \lambda} \neq \frac{l}{\lambda}$. (In fact, it is
much smaller than $l/\lambda$).
The reason for this is obvious: the $\delta \lambda$'s from the $l/ \lambda$
parts in $l$ do {\em not} add coherently.

Next we have to find how much the speed of light fluctuates. We proceed as
follows:
Just as there are uncertainties in spacetime measurements, there are
also uncertainties in energy-momentum measurements due to
spacetime foam effects. Thus there is a limit to how accurately we
can measure and know the energy and momentum of
a system~\cite{ng94,ng95}. 
Imagine sending
a particle of momentum $p$ to probe a certain structure of spatial extent
$l$ so that $p \sim \hbar/l$.
It follows that $\delta p \sim (\hbar/l^2) \delta l$. Spacetime
fluctuations
$\delta l \gtrsim l (l_P/l)^{\alpha}$ can
now be used to give
\begin{equation}
\delta p \sim p \left(\frac{p}{m_P c}\right)^{\alpha}.
\end{equation}
The corresponding statement for energy uncertainties is
\begin{equation}
\delta E \sim E \left(\frac{E}{E_P}\right)^{\alpha}.
\end{equation}
Here $m_P$ and $E_P = m_P c^2$ denote the Planck mass and energy 
respectively.
These energy-momentum uncertainties modify
the dispersion relation for the photons to read:
\begin{equation}
E^2 \simeq c^2p^2 + \epsilon E^2 \left(\frac{E}{E_P}\right)^{\alpha},
\end{equation}
where $\epsilon \sim \pm 1$. Thus 
the speed of light 
$v = \frac{\partial E}{\partial p}$ fluctuates by 
\begin{equation}
\delta v \sim 2 \epsilon c (E/E_P)^{\alpha}.
\end{equation}
We emphasize that all the fluctuations take on $\pm$ sign with equal
probability (like a Gaussian distribution about $c$).

It follows from the above discussion that
for photons emitted simultaneously from a distant
source coming towards our detector, we would expect an energy-dependent
spread in their arrival times. To maximize the spread in arrival
times, we should look for energetic photons from distant sources.
So the idea is to look for a noticeable spread in arrival
times for such high energy gamma rays from distant gamma ray
bursts. This proposal was first made by~\cite{ame98} in another context.

To underscore the importance of using the correct cumulative factor
to estimate the spacetime foam effect, let us first proceed in a naive
manner.
At first sight, the fluctuating speed of light would {\em seem} to
yield an energy-dependent spread
in the arrival times of photons of energy $E$ given by
$\delta t \sim \delta v (l/c^2) \sim t (E/E_P)^{\alpha}$,
where $t = l/c$ is the average overall time of travel from the photon source
(distance $l$ away).
But these results for the spread of arrival times of photons are
{\em not} correct, because we have inadvertently
used $l/\lambda \sim Et/\hbar$ as the cumulative factor instead of the
correct factor $(l/\lambda)^{1 - \alpha} \sim (Et/\hbar)^{1 - \alpha}$. Using the
correct cumulative factor, we get a much smaller 
\begin{equation}
\delta t \sim t^{1 - \alpha} t_P^{\alpha} \sim \delta l / c
\end{equation}
for the spread in arrival time of the
photons~\cite{ng08}, independent of energy $E$ (or photon wavelength, $\lambda$). 
Here $t_P \sim 10^{-44}$ sec 
is the 
minuscule Planck time.
Thus the result is that the time-of-flight differences
increase only with the $(1 - \alpha)$-power of the
average overall time of travel $t = l/c$ from the gamma ray bursts to our
detector, leading to a time spread too small to be detectable (except for the 
uninteresting range of $\alpha$ close to 0).

The new Fermi Gamma-ray Space Telescope results~\cite{abdo09}
of $\delta t \lesssim 1$~s for $t \sim 7$ billion years, rule out only 
spacetime foam models with $\alpha \lesssim 0.3$.  We note that
the result $\delta t \gtrsim t^{1 - \alpha} t_P^{\alpha}$
is to be expected if we recall that the spread of 
arrival times can be traced to uncertainty in the distance that the 
photons have travelled from the distant source to our telescopes.
In other words, gamma ray arrival time dispersions $\sim 1$~s from sources
at distances $\sim 7 \times 10^9$~ly only reject spacetime foam models in which the 
effects almost add coherently ($0 \leq \alpha < 0.3$) but do not test the
major models discussed above (i.e., $\alpha = 1/2, 2/3, 1$). For example the
holographic model predicts an energy independent dispersion of arrival times
$\sim 2.5 \times 10^{-24}$~s. 

Finally, in the following section on spacetime foam induced
uncertainties in wave vector direction we show angular uncertainties
in the wave vector direction are not the result of random multiple scattering
events which could, in principle, lead to much larger uncertainty
in arrival times.

\section{Angular Fluctuations: Measurement}
\label{sec-meas}

At the outset it must be re-emphasized that the test for spacetime foam effects
is a null test.   A theoretical model for spacetime foam is disproved if images
of a distant point source do not exhibit the blurring predicted by theoretical
spacetime foam models.

As discussed in the Introduction, the fundamental uncertainties caused by
spacetime foam are spatial, not angular.  Strictly speaking, the models specify
the uncertainty $\pm \delta l$, in distance between a source and observer along
the line of sight.  This is because $\delta l$ is defined by the uncertainty in
the distance measured by light travel times.   Straightforward extensions of the
models hypothesize that $\delta l$ is isotropic in the sense that the
uncertainty perpendicular to the line of sight is equal to the line of sight
uncertainty (cf. CNvD, 2006).   Of course, there is also a corresponding
uncertainty in the transit time for light from source to observer,  $\delta t
\sim \delta l/c$, which was discussed in section III.   Furthermore, since the
globally averaged wavefront is effectively spherical it implies that globally
averaged photon trajectories deviate from the direct line of sight by an angle
less than or equal to $\delta l/l$ .  A direct consequence of this limitation on
the globally averaged angular deflection of photon trajectories is that the
theoretically expected blurring of distant images due to spacetime foam is not
the result of a random walk of small angle photon scatterings along the line of
sight, since the uncertainties in the derived directions of the local wave
vectors must result in the same spatial uncertainty, $\delta l$ (no matter how
many wave crests pass the observer's location).  For example, in the "thin
screen approximation", the accumulated transverse path of multiply scattered
photons is approximated as $(\delta \psi)l >> \delta l$.  This would lead to
expected time lags, $\delta \psi (l/c) >> \delta l/c$, in conflict with the
basic premises for space time foam models.   

The reasons why a straight-forward application of standard small angle
scattering theory (e.g., gravitational micro-lensing) is not applicable are
two-fold:

{\noindent (1) the "scattering centers" for spacetime foam would be virtual
Planck masses that exist for only a Planck time ($\sim 10^{-44} $ sec) which is
much less than the period ($\sim 10^{-28}$ sec) of even the highest energy (TeV)
detectable  gamma rays.

(2) the "density" of such virtual scattering centers can only be inferred from
nonlocal theories such as the holographic principle. .    

However, according to theory, although the wavefront remains approximately
spherical its precise location and structure is best characterized as a "cloud"
of uncertainty whose scale is $\delta l$.  These uncertainties are not realized as
fluctuations until physical measurements are performed.   This is consistent
with the quantum physics of these models with their reliance on the quantum
Uncertainty Principle.  }

As the light wave encounters a "detector" (e.g. a telescope) the spatial
uncertainty shows up as a random series of fluctuations of scale $\leq (\pm) 
\delta x$ along and perpendicular to the line of sight causing fluctuations (i.e.
"jitter") in the apparent angular direction of the wave vector recorded by the
detector.  However, the cloud of uncertainty associated with these spacetime foam
models requires that the spatial uncertainties in the wave, from crest to crest,
do not continuously connect.   This cloud of uncertainties is chaotic and
fluctuations are not correlated from crest to crest.  That is, you cannot trace
caustics from one crest to another.  

Furthermore, it must be understood that a necessary condition for measurement
of  angular blurring requires that $\delta l << \lambda$, which corresponds to
the limit for quantum coherence.  As the electromagnetic wave from the
extragalactic source propagates, the phase uncertainty $\Delta \phi \sim 2 \pi
\delta l / \lambda$ gets larger.  Once $\Delta \phi \sim 2 \pi$ (corresponding
to $\delta l \sim \lambda$) quantum coherence is lost.  At this point, while the
amplitude of the electromagnetic wave remains intact, all phase information is
lost.  The distance at which spacetime foam induced de-coherence sets in depends
on the model.   For the "$\alpha$-models," coherence is lost sooner when
$\alpha$ is smaller.

In the realm where imaging is possible ($\delta l << \lambda$), detecting
spacetime foam effects via "blurring" must involve a form of  triangulation by
comparing $\delta l$ with some well-defined localized length scale, $B$, that
would allow for the measurement of an angular uncertainty defined by the
dimensionless ratio,  $\delta \psi \sim \delta l/B$.  Possible length scales
defining $B$ that may be relevant are:  1)  the diameter of the telescope, $D$,
or 2)  the wavelength of the radiation, $\lambda$.  In both cases the angular
uncertainty would represent the theoretically predicted apparent broadening of
the image of a point source of radiation caused by spacetime foam as recorded by
a telescope plus detector

GEOMETRIC OPTICS:

If the telescope/detector does not record phase information, then angular
broadening at the focal plane will be determined by ray-tracing. Therefore, the
spatial uncertainty, $\delta l$, in the location of a ray on the primary
(whether it is a lens or a mirror), will result in an angular uncertainty, at
the focal plane of,

\begin{equation}
\delta \psi |_{Geometric ~ Optics} \sim \delta \psi |_{min} \sim
2\delta l / D << \lambda/D.
\label{psi2}
\end{equation}

(Of course the focal ratio of the telescope also plays a role in the above
equation, but the key parameter is the size of the primary.)  When combined with
the requirement for quantum coherence discussed above (i.e., $\delta l <<
\lambda$), eq. (7)
virtually guarantees that spacetime foam effects will be undetectable by
techniques that are phase insensitive (i.e. geometric optics).

WAVE OPTICS:

Clearly the largest measurable angular effect would be associated with using the
shortest length standard, i.e. the wavelength.   As stated previously, the only
way to accomplish such a measurement is via interference techniques such as
diffraction limited imaging and/or interferometry.  In this case, since 
$\delta l = N l^{(1-\alpha)} l_P^\alpha$,  
the angular uncertainty (i.e. the blurring angle)  is, 

\begin{equation}
\delta \psi |_{Wave ~ Optics}=\delta \psi_{max}=\delta l/\lambda =
N (l/\lambda)^{(1-\alpha)}(l_P/\lambda)^\alpha.
\label{psi3}
\end{equation}

This relation is identical in form to the relation defining $\delta l$, with
dimensionless ratios, $(l,l_P)/\lambda$,
replacing the previously defined distances.  
Note the so-called "accumulation factor, $(l/\lambda)^{(1-\alpha)}$, 
plays the same role as
discussed previously in the growth of the distance uncertainty and the time
uncertainty discussed in section III. The magnitude of $\delta \psi$ as given in
Eq. (8) is consistent with our assumption of isotropic fluctuations which
implies comparable sizes of the wave-vector fluctuations perpendicular ($\delta
k_{perp}$) to and along ($\delta k_{par}$) the line of sight (cf. CNvD, 2006). 
To wit, the size of the angular spread $\delta \psi$ is given by $\Sigma \delta
k_{perp} / k$ where $\Sigma$ stands for summation over the $l/\lambda$ intervals
each of length $\lambda$.  Since $\delta k_{perp} \sim \delta k_{par}$ according
to the assumption of isotropic fluctuations and $k_{par} \approx k = 2 \pi /
\lambda$, we have $\delta \psi \sim \Sigma \delta k / k = \Sigma \delta \lambda
/ \lambda \sim \delta l / \lambda.$

Thus, interference techniques offer the best (perhaps only)  method for testing
for the presence of spacetime foam because the observed wavelength offers the
shortest line of sight baseline calibrator for measuring the local angular
uncertainty in wave vector direction and interference is the only way to exploit
the wavelength as a calibrator.  That is, the expected blurring of diffraction
limited images is a consequence of interference effects resulting from wavefront
"distortions" (or "measurement errors") corresponding  to ratios of positional
uncertainties of order $\pm \delta l$ along, and perpendicular, to the line of
sight  and the corresponding angular uncertainties $\pm \delta l / \lambda$.  In
particular, the angular uncertainty corresponds to the ''wavefront errors" that
are the basis for calculating Strehl ratios in wave-optics.  Since the Strehl
ratio is designed to provide a quantitative measure of the departures from
diffraction limited optics, the technique is well suited for providing the
proposed null test of spacetime foam models.  Simply put, if distant point
sources do not show reductions in Strehl ratios predicted by theoretical models,
the model must be rejected.  The application of this technique is discussed in
detail in section VI below.  

We end this section with a summary of what we expect a telescope would observe
at different locations.  As the electromagnetic wave from the extragalactic
source propagates,  the phase uncertainty $\Delta \phi \sim  2 \pi \delta l /
\lambda$ gets  larger.  So long as $\Delta \phi <<  \lambda /D $ (in other
words, so  long as $\delta l << \lambda^2 / D$), normal interference pattern is 
observed.  Farther from the source, as $\Delta \phi \sim \lambda / D$  (i.e.,
$\delta l \sim \lambda^2/D$), we expect the interference pattern  to be replaced
by a halo structure.  But once $\Delta \phi $ reaches  $\sim 2 \pi$
(corresponding to $\delta l \sim \lambda$) , quantum coherence  is lost, and the
images will become all blurred.  At this point, while  the amplitude of the
electromagnetic wave remains intact,  all phase  information is lost.  How far 
from the source these different stages (from interference pattern to  halo
structure to all blurred images) are encountered, of course,  depends on the
different spacetime foam models. For example, for the AGN PKS 1413+135
\cite{per02}, the random-walk model predicts that $\Delta \phi \sim 10 \times 2
\pi$, while the holographic model yields $\Delta \phi \sim 10^{-9} \times 2 \pi$
for $\lambda \approx 1.6 \mu$ m \cite{chr06}.  Thus, according to the
random-walk model, $\Delta \phi$ has reached the stage beyond the quantum
coherence limit and hence the image should be all blurred.  However, even
supplemented with the assumption of isotropic fluctuations (viz., $\delta
k_{perp} \sim \delta k_{par}$) which henceforth we adopt, the holographic
model predicts that a normal interference pattern is observed at the HST (with 
$D \approx 2.4$m) for $\lambda \approx 1.6 \mu$m.  On the other hand, if $D$
were about 3 orders of magnitude larger, according to the holographic model, a
halo structure would be observed.

\section{Testing Spacetime Foam Models Using Astronomical Observations}
\label{sec-obstest}
With the above as background, we are now prepared to discuss the issue of
testing spacetime foam models. This subject has been discussed previously
by CNvD as well as~\cite{ste07} and~\cite{rag03}. 

\begin{figure}%[htbf]
\centering
{\includegraphics[width=9.5cm]{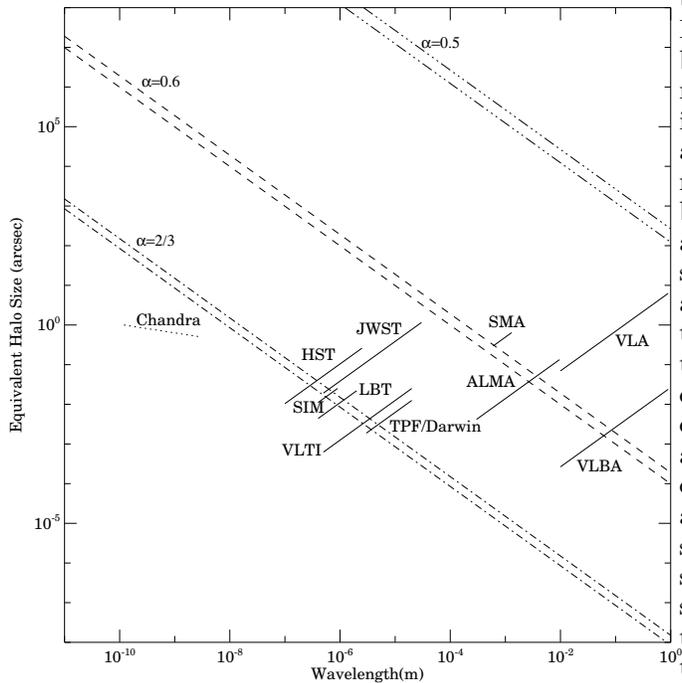}}
\caption{\label{fig:det} The detectability of various models of foamy spacetime 
with existing and planned telescopes. Diagonal tracks are shown for three 
models of foamy spacetime, namely $\alpha=0.5, 0.6, 2/3$, for $z=1$ and $4$ 
(respectively the lower and upper tracks for each model) and $N=1.9$.
Also shown are the observing ranges and theoretical resolution limits 
(i.e., PSF size) for a wide variety of telescopes, both current and planned. 
The tracks for each telescope represent the diffraction limit with the 
exception of the dotted track for {\it Chandra}, whose complicated 
PSF is NOT diffraction limited (see
section~\ref{sec-obstest}). 
As such, they represent the detectability limit for these observations, in the 
case of a perfect telescope of that size. 
Importantly, note that in practical 
terms a given telescope may not be able to probe quite as far as pictured, 
as in the final analysis the region probed is also a function of Strehl ratio 
(see section~\ref{sec-HST}). }
\end{figure}

CNvD described a more flexible way to use astronomical
observations, which has been summarized and expanded upon in the preceding
sections. As we note in the Introduction, for point sources of low redshift
$z$, it is not so critical which distance expressions for the point sources
one should use. But for high redshift objects, it
does make a difference. We have already argued that the appropriate
cosmological distance to use is the total line-of-sight comoving
distance~\cite{hog00} given by
\begin{equation}
D_C(z) = D_H \int_0^{z} dz'/E(z'),
\label{eq1}
\end{equation}
where
\begin{equation}
E(z) = \sqrt{\Omega_M (1+z)^3 + \Omega_k (1 + z)^2 + \Omega_{\Lambda}},
\label{eq2}
\end{equation}
with $D_H = c/H_0$ being the Hubble distance, $\Omega_M, \Omega_k$ and
$\Omega_{\Lambda}$ being the (fractional) density parameter associated with
matter, curvature and the cosmological constant respectively. Consistent
with the latest WMAP + CMB data, we will use
$\Omega_M = 0.25, \Omega_{\Lambda} = 0.75$ and
$\Omega_k = 0$, and for the Hubble distance we will use
$D_H = 1.3 \times 10^{26}$ meters.
In terms of the comoving distance, for the various models of spacetime foam
(parametrized by $\alpha$), the equivalent halo size is given by
\begin{equation}
%\Delta \phi /2 \pi \approx D_C^{1-\alpha} l_P^{\alpha}/\lambda_e,
\delta \psi = N (1 - \alpha) l_P^{\alpha} D_H^{1 - \alpha} I(z,
\alpha)/ \lambda_o, 
\label{eq3}
\end{equation}
with
\begin{equation}
I(z,\alpha) = \int_0^z dz' (1+z')/(E(z')) \left(\int_0^{z'}
dz''/E(z'')\right)^{-\alpha},
\label{eq4}
\end{equation}
where the factor $(1 + z')$ in the integral corrects the observed wavelength
$\lambda_o$, back to the wavelength
$\lambda (z')$ at redshift $z'$. That is
$\lambda (z') = \lambda_o/(1 + z')$.
We have used these results to produce Figure~\ref{fig:det}, which shows the
predictions made for three different values of $\alpha$, i.e.
$\alpha = 2/3, 0.6, 1/2$ respectively, for the size of observed haloes
produced by accumulated phase
dispersion for a source at two redshifts, respectively $z=4$ and $z=1$.
This graph plots wavelengths from hard X-rays (equivalent photon energy
$\sim 100$ keV) down to radio waves.
The labeled arrows pointing upward to the right delineate the diffraction
limited response of various existing or proposed telescopes. That is, with the 
exception of {\it Chandra}, the arrows plot $\lambda_o/D$, where D is the 
aperture/baseline length and the length of the arrow is determined by the 
telescope's wavelength limits. The shallower slope of {\it Chandra's} results 
from
its grazing-incident optics and its complicated point spread function.

As $\delta \psi$ depends nontrivially on the redshift of the source, a 
remark on the effect of the uncertainties in the determination of the sources
is in ordered.  We have calculated the ratio $\delta I/I$ for a variety of 
redshifts and uncertainties.  Some sample values are given in Table
~\ref{tab:delI}.  As can be seen, the effect is small, unless the uncertainty
in the redshift, $\delta z$, grows significantly larger than 0.1, as would be 
the case for photometric redshift determinations \cite{Mob07,Wu10}

Steinbring \cite{ste07} showed a plot of a somewhat similar nature based, 
however, on luminosity distance rather than the co-moving distance 
(his Figure 1). The main differences here are corrected curves for 
$\alpha=2/3$ and $1/2$, since~\cite{ste07} also did not show the
curves yielded by CNvD. 

We note that if the arrow representing a telescope's diffraction limited response
lies {\em below} the halo size curve for a given $\alpha$, that model may be 
excluded by observations. For example, as mentioned previously the 
random walk model ($\alpha = 1/2$) is clearly excluded by \HST observations, 
as well as radio observations by the NRAO Very Long Baseline Array (VLBA).
The latter surveys find numerous examples of unresolved sources
at cosmological redshifts~\cite{kel98,kel04}. We
also note that \HST observations at short wavelengths ($\sim 10^{-6} - 10^{-7}$ m)
appear to be capable of testing the holographic model ($\alpha = 2/3$) and 
certainly may exclude $\alpha = 0.6$.

Before delving further into the observational record to discuss what models of
spacetime foam are excluded by the current data, it is useful to make a few
comments. First of all, what types of observations and objects would be
useful for constraining or rejecting spacetime foam models?
In principle, all one needs are high-resolution
imaging observations of a source that
is unresolved and at high redshift.
However, as noted previously, for a variety of practical
reasons the source must be reasonably bright. This is for two reasons.
First, one needs to be able to reliably distinguish a halo produced by
dispersions induced by quantum fluctuations in spacetime, from a faint
``halo'' that might be due to intrinsic structure in the source. For
example, in the optical, every bright extragalactic point source, be it a
supernova, QSO, GRB or otherwise, is resident in a galaxy, which could very
easily mimic or disguise a halo. Thus
one wants to be able to achieve large dynamic ranges (at least hundreds or
thousands to one) on the image with a reasonable amount of integration time.
This host galaxy can also scatter light from the AGN, so that there may still 
be some halo left in the image (e.g., \cite{ogle07}).  However, the scattered
light is polarized, so that there is a clear way to discriminate  
the effects of spacetime foam from scattering effects.
Secondly, a number of observing methods, particularly optical
interferometry, require much brighter sources in order to be able to find
fringes in real time. For example, the design of the VLTI requires sources
in the IR K-band brighter than 14th magnitude. Folding these constraints in, for
non-interferometric observations in the optical, any bright point source
will do, be it a QSO or GRB. Unfortunately, SN Ia or SN II cannot be used, 
as they are comparable in luminosity to the host galaxy so they
will not allow large enough signal to noise to be reached (i.e., we will not
be able to distinguish structure in the host galaxy from broadening due to
spacetime foam). However, for
optical interferometers such as the VLTI or (in the future) SIM or
Darwin/TPF, only the brightest Blazars observed in
outburst, or the prompt flash of a powerful GRB will do. 

In the radio, any bright QSO or BL Lac object will suffice, and fortunately 
there are large numbers of these. A large number have been found to 
have unresolved sources, as seen with both the NRAO VLA and 
VLBA~\cite{kel98,kel04}. These observations confirm at radio 
wavelengths that the random walk model can be ruled out. None of 
these observations have sufficient resolution to test the holographic 
model, however.

Ideally, one wants to observe with any telescope that will yield the
smallest $\lambda / D$ ratio as the diffraction limit would then dictate 
that one could set the smallest limit on the size of any halo corresponding 
to the wavelength at which a given spacetime foam model would predict 
the largest halo size.
However, for one of the telescopes shown on Figure 1, namely the 
{\it Chandra} X-ray Observatory, this is not the case.  Unlike the radio, IR 
and optical telescopes shown there, {\it Chandra}'s optical design is 
governed by the short wavelengths (comparable to inter-atom spacings) 
of X-ray photons, which make ordinary normal-incidence optics impractical. 
Instead, {\it Chandra} uses grazing incidence mirrors in a 
nested~\cite{wol52a,wol52b} design.  
In this design, it is the nesting of the mirrors that focusses the light on the 
imaging plane.  As a result, the PSF is not diffraction limited, and the 
telescope cannot sense the wavefront coherently.  We have plotted {\it Chandra}'s
angular resolution in Figure 1 using parameters from the {\it Chandra Proposer's 
Observatory Guide}, using values corresponding to the 75\% level in the 
encircled energy function (see specifically Figures 4.6 and 4.7).  
However, in Figure 1 we show the {\it Chandra} line as dotted.  
This is because its inability to sense the wavefront coherently makes it 
unsuitable for testing models of spacetime foam because the blurring 
of the images is a product of interference effects rather than photon 
scattering, an effect which non-diffraction-limited optics are not 
sensitive to (see section~\ref{sec-meas}).

%However, for one of the telescopes shown on Figure~\ref{fig:det} this is not the 
%case. The exception is the {\it Chandra} X-ray Observatory 
%(dotted line in Figure~\ref{fig:det}), for which the angular
%resolution is nearly constant and in fact is slightly larger at shorter
%wavelengths (higher X-ray energies), contrary to what one would expect due
%to simple diffraction. The reason behind this is the design of {\it
%Chandra}, which uses grazing incidence mirrors in a 
%nested~\cite{wol52a,wol52b} design. In this design, it is
%the nesting of the mirrors that focusses the light on the image plane, and
%as described in the {\it Chandra Proposer's Observatory Guide} (hereafter
%POG, specifically section 4.2.3), this focussing is slightly better at
%longer wavelengths. The result is a PSF that varies in size by roughly a
%factor 2 across the {\it Chandra} band of 0.3-10 keV. The angular size
%plotted in Figure~\ref{fig:det} corresponds to roughly the 75\% level in the 
%encircled energy function (see POG, Figures 4.6 and 4.7).

Of course one would also like to be able to obtain constraints in as many
bands as possible, since %The reason is that finding the
%one is trying to distinguish a halo
%due to dispersion from spacetime foam from intrinsic structure. mapping
finding the halo size as a function of frequency, in principle, allows one to
search for the characteristic $\lambda^{-1}$ dependence in the size of the
spacetime foam scattering disk (discussed in section~\ref{sec-meas}). 
%Since the particulars of that structure may be different in each band, having
%constraints in several bands is the ideal situation. 
Constraints obtained in only one band are susceptible to bias by structure
%constraints due to faint source and/or
intrinsic to the source (host galaxy or extended emission line regions, for example) or to the PSF of that band
% modulo any constraints due to the nature of the PSF in that band. %??? LOTS OF CONSTRAINTS!!!

%Finally, the need to reliably distinguish between intrinsic and
%foam-produced structures also means that one needs relatively easily
%distinguished markers of an unresolved source. In most bands and with most
%telescopes this is relatively easy to do; however, with the {\it Chandra
%X-ray Observatory} it is somewhat more difficult. This is because the wings
%of the {\it Chandra} PSF are much broader than the core, as an inescapable
%result of the telescope's design. Specifically, the {\it Chandra} POG,
%Figure 4.7 shows that the 90\% level in the EEF is at an angular scale 2-10
%times greater than that of the 75\% value plotted in our Figure~\ref{fig:det}. This makes
%distinguishing between intrinsic and foam-produced structure rather more
%difficult. In fact, to be able to do this reliably in the X-rays one would
%need to be able to obtain a high signal to noise (hundreds to one) in a very
%small energy range so that any energy dependence could be minimized. Thus
%despite the fact that a superficial look at Figure~\ref{fig:det} would suggest that
%{\it Chandra} is more sensitive to foam than any other telescope currently
%built or planned, it is not certain that this can be done with the current
%archive although we will pursue this subject further
%in a followup paper.

With the above provisos, then, we can now discuss the constraints that can
currently be put on spacetime foam models. It is apparent that with current
telescopes, all spacetime models with $\alpha \leq 0.6$ are firmly excluded
by the combined archives of telescopes in the optical, radio and X-ray where
unresolved sources have, at some level (see below) been observed at every
redshift up to at least $z=5$. With \HST %and {\it Chandra}
observations, this constraint can be significantly tightened. % to $\alpha>0.65$.
Thus the random walk ($\alpha=1/2$) model is securely excluded at very 
high confidence, as in fact are all models with $\alpha \leq 0.65$. 
The Holographic ($\alpha=2/3$) model is very close to
being tested and will enter that regime once the {\it VLTI} is fully
operational, and may be tested by \HST in the ultraviolet.

%V. {\bf Testing Spacetime Foam Models - The HST Quasar Archive}
\section{Testing Spacetime Foam Models - The \HST Quasar Archive}
\label{sec-HST}
\begin{figure*}
\centering
{\includegraphics[width=7.3cm]{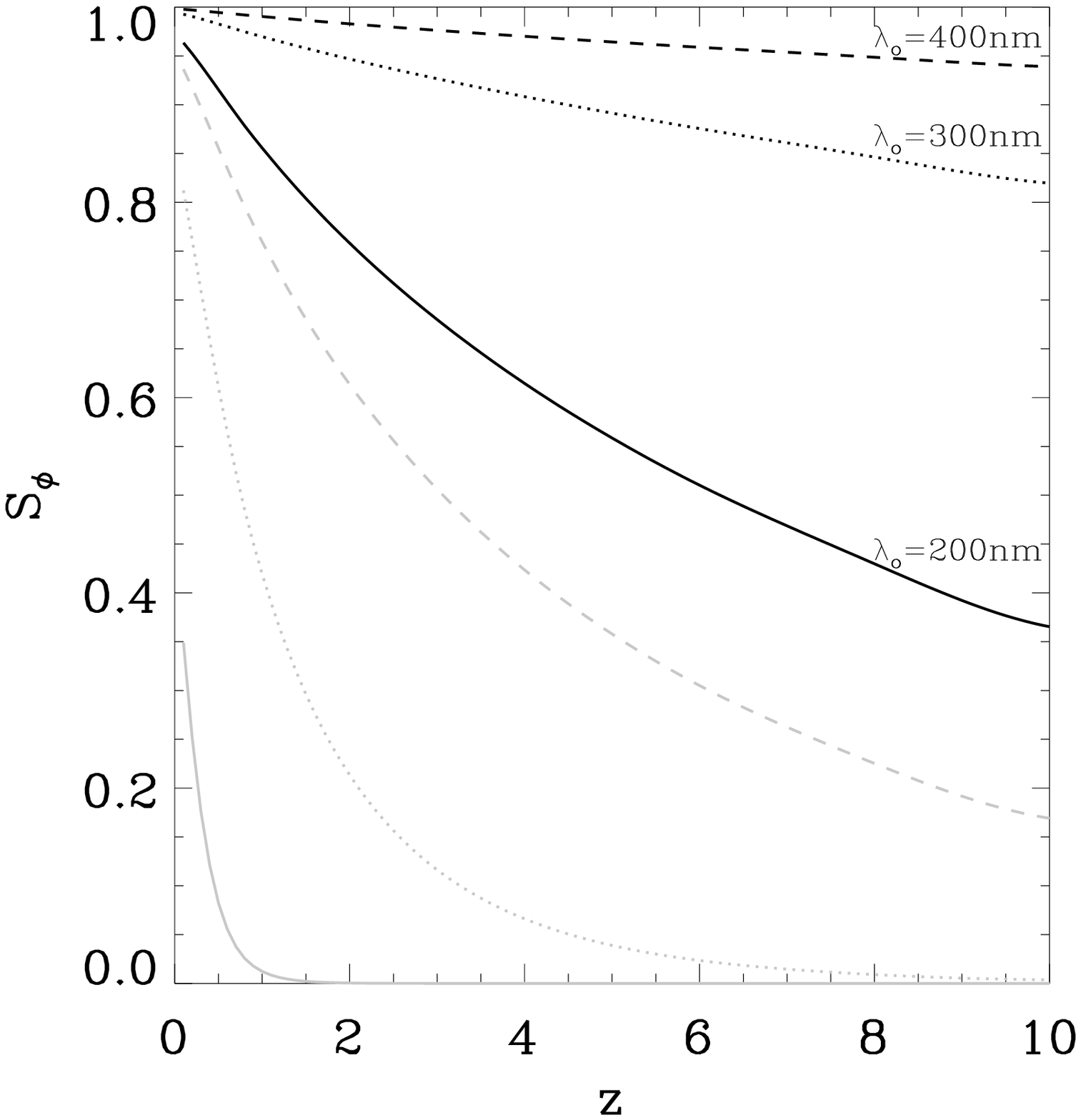}}
{\includegraphics[width=7.3cm]{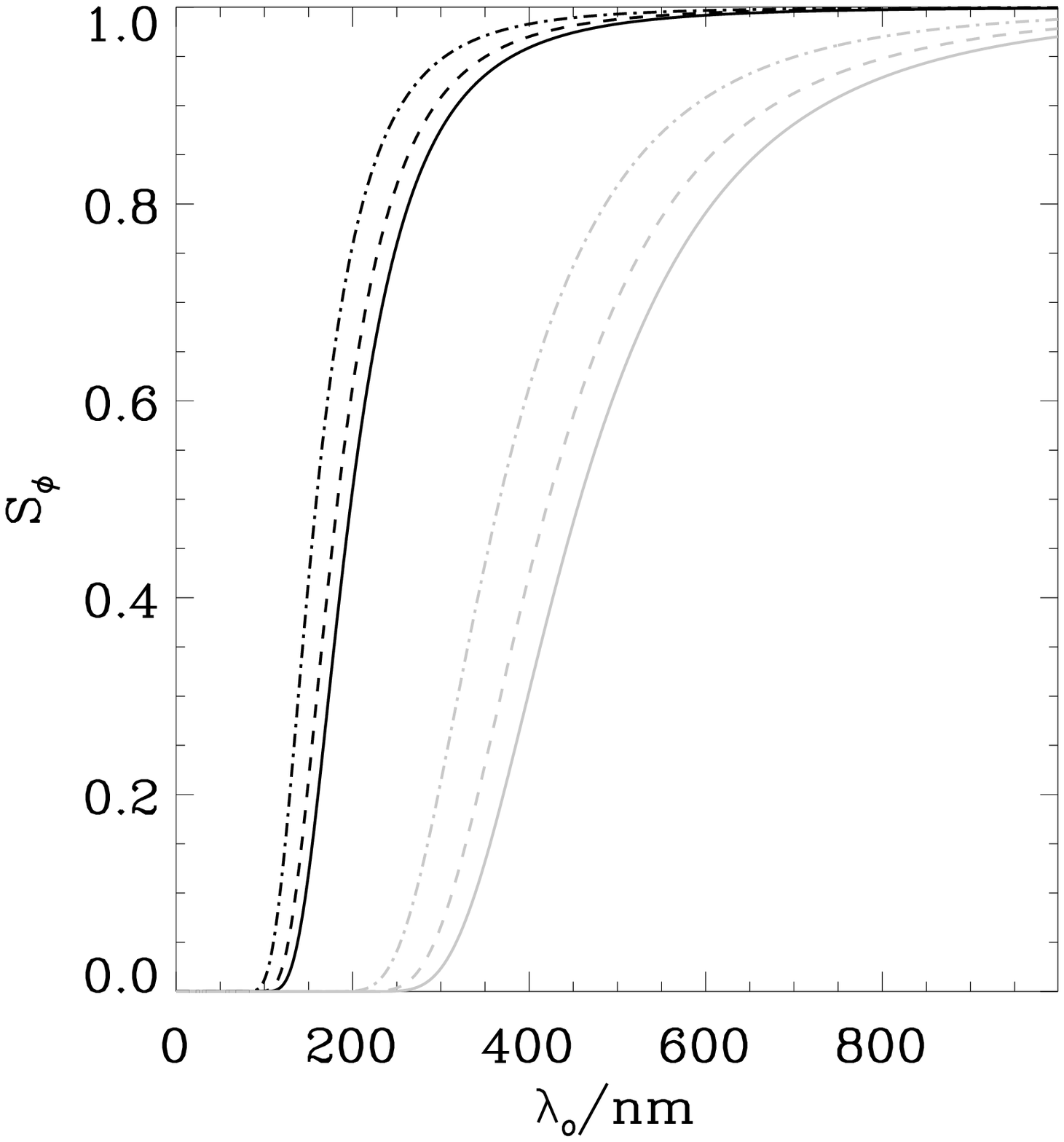}}
\caption{\label{fig:SRT} Expected Strehl ratio with redshift (left) and 
wavelength (right) for two spacetime foam cases:
$\alpha=2/3,~N=1.9$ (black lines);
$\alpha=0.655,~N=1.9$ (grey lines).
In the left-hand figure we illustrate the change in Strehl ratio expected for
a point source with varying distance, $z$, for three fixed observed 
wavelengths ($\lambda_o=200, 300$ and $400$~nm as marked).
In the right-hand figure we show the change in Strehl ratio expected for 
a point source at various three fixed distances: 
$z=6$ (solid line), 
$z=4$ (dashed) and 
$z=2$ (dot dashed). 
}
\end{figure*}

Because the Strehl ratio is defined as the ratio of the observed image peak
to the peak diffraction spike of an unabberated telescope, it provides a 
useful way to assess the effects of spacetime foam.
As discussed above, spacetime foam causes wavefront ``errors''
in the images of distant point sources. The effect is similar to 
atmospheric ``seeing'' and, in fact, we follow the general analysis of 
Sandler et al. \cite{sandler94} for the degradation of Strehl ratios resulting from 
wavefront errors. Since spacetime foam induced wavefront errors are not 
spatially correlated, a simple approximation for the degraded Strehl ratio is
\cite{sandler94}
\begin{equation}
S_\mathrm{Obs} = S_M \exp [-(\sigma_I^2 + \sigma_{\phi}^2)]
\label{eq5}
\end{equation}
where $S_M \leq 1$ represents a degradation of the observed Strehl ratio
due to masking effects (which may or may not be spatially correlated as 
discussed above),
$\sigma_I$ represents uncorrelated wavefront errors induced by the instrumentation
(i.e., telescope plus instruments) and $\sigma_{\phi}$ represents uncorrelated wavefront
``errors'' induced by spacetime foam. Both of these
dispersions are expressed in units of the telescope's diffraction limit, 
$\lambda / D$. For convenience we then define the spacetime foam degraded
Strehl $S_{\phi}$ as:
\begin{equation}
S_{\phi} = \exp (-\sigma_{\phi}^2)
\label{strehl}
\end{equation}
where $\sigma_{\phi}$ is simply the angular dispersion of the wavefront vector
due to external effects
(i.e., $\delta \psi$ in Eq.~(\ref{eq3})) divided by 
$\lambda / D$, 
\begin{equation}
\sigma_{\phi} = N (1 - \alpha) l_P^{\alpha} D_H^{1 - \alpha} I(z,
\alpha) D/ \lambda_o^2.
\label{sigma}
\end{equation}
($S_{\phi} \sim 1$, of course, corresponds to no image degradation.)
This approximation, of course, breaks down when $\sigma_{\phi} \sim 1$, i.e.,
when the wave front angular dispersion is comparable to the telescope's 
resolution. A fully parametrized version of the resultant Strehl ratio then is
\begin{equation}
S_{\phi} = \exp \left(- N^2 (1 - \alpha)^2 l_P^{2\alpha} D_H^{2(1 - \alpha)} 
I^2 (z,\alpha) D^2/ \lambda_o^4 \right).
\label{eq:SR}
\end{equation}
A few comments concerning parameter sensitivity of these models for Strehl ratio 
are in order:
\begin{enumerate}
\item The model Strehl ratio is strongly sensitive to wavelength.
\item Although we have managed (section~\ref{sec-dfe}) to narrow the range of possible values 
for the constant, $N$, in the holographic model 
the remaining uncertainty in this parameter
($0.97 < N < 1.9$) leads to a factor of 3.4 in the argument of the Strehl ratio
exponential.
\item The instrumental contribution, $S(\sigma_I)$ can be measured using a true 
point source, for example a star.
\item The model Strehl ratio is also exquisitely sensitive to the value of 
$\alpha$.

\end{enumerate}
Figure~\ref{fig:SRT} illustrates the sensitivity to degradation of $S_{\phi}$ from 
spacetime foam as a function of redshift and of wavelength for $\alpha$ 
values of $2/3$ (in black) and $0.655$ (in grey) -- assuming $N=1.9$ in each case. 
In the left-hand figure we illustrate the change in $S_{\phi}$ expected for a point 
source with varying distance for three fixed wavelength values. 
In the right-hand figure we show the change in $S_{\phi}$ expected for a point 
source with varying wavelength at three fixed distances.
We can see that \HST is just able to probe the $\alpha=2/3$ case at the very 
shortest (UV) wavelengths in its spectral range. 
%The strong wavelength dependence is illustrated amply in Figure~\ref{fig:SRT}: 
%Figure~\ref{fig:SRT} also illustrates the wavelength-$\alpha$ sensitivity of the spacetime
%foam induced degradation of $S_{\phi}$ with increasing redshift for $N = 1.8$. 
We note that $N = 0.97$ shifts the curves in the left-hand figure 
to the right (i.e., increasing redshift). We also note that
shorter wavelength is a greater imperative than increased distance for 
improving sensitivity to $\alpha$.
However, optical measurements may rule out lower values of $\alpha$.
Clearly the realm for testing the holographic model lies in
redshifts $z \gg 1$. Notice, also, that even a modestly smaller value of 
$\alpha = .64$ clearly can be tested at redshifts more than an order of magnitude
smaller.

To test the spacetime foam models using Strehl ratios, we explored the \HST 
archive (http://archive.stsci.edu/hst/search.php) for observations
of high-redshift quasars, using StarView to cross reference suitable
catalogues with the archive database. We explored the Sloan Digital 
Sky Survey (SDSS) DR5 quasar catalogue~\cite{schneider+07} and 
a catalogue of the highest redshift SDSS quasars~\cite{richards+04} 
in addition to the quasar catalogue of~\cite{VCV10} -- hereafter VCV.
In Table~\ref{tab:hst} we present notable high-redshift quasars observed in
the optical, NIR and UV using \HST. 
We prioritize those observations that maximize the value of 
$I(z,\alpha)/\lambda_o$ (see Eq.~(\ref{eq4})).

The current record for the highest redshift quasar is held by the CFHT 
quasar survey: CFHQS~J2329-0301 at $z=6.43$. This has not yet been 
observed with \HST. The highest redshift quasars with 
\HST pointings all come from the SDSS, including 
SDSS~J1030+0524 at $z=6.28$.
and SDSS~J1148+5251 at $z=6.42$
The former was observed, along with other high-$z$ SDSS quasars, using the 
{\em Advanced Camera for Surveys (ACS) High Resolution Channel (HRC)} 
at $\lambda_o\sim850$nm by~\cite{richards+04}. 
Blurring of these images has been studied inconclusively~\cite{ste07}.
However, the observations are all at red wavelengths 
(F775W and F850LP), which impinges on their sensitivity to spacetime foam, 
especially the $\alpha=2/3$ case (see Eq.~\ref{eq:SR}, and Figure~\ref{fig:SRT}).
The latter was observed through two linear ramp filters by~\cite{white+05} 
at $\sim720$ and $\sim900$~nm, and by~\cite{stia03} 
in F850LP and F775W.
Among other notable examples is the HDF-S QSO at
$z=2.24$~\cite{savaglio98}, which was observed as part of the
multi-wavelength coverage of the deep fields using the {\em Space Telescope
Imaging Spectrograph (STIS)} in the ultraviolet. 
%at a rest wavelength of $\lambda_r=66$~nm. 
While offering the big advantage of shorter wavelengths,
the STIS data is unsuitable for the present purpose due to the extreme
distortion of the PSF, which appears triangular with no discernible Airy
ring. 
%A sample of 10 quasars at $z\approx2$ exists in the work of~\cite{kukula+01}, 
%but these were observed at near-infrared wavelengths using the 
%{\em Near-Infrared
%Camera and Multi-Object Spectrometer (NICMOS)}
%(corresponding to $\lambda_r\approx$ 550~nm).

We were unable to find any short wavelength ($\lambda_o<500$~nm) 
images of high redshift ($z>4$) quasars by cross-referencing with SDSS or 
VCV quasar catalogues. These short wavelength observations are critical 
because they represent the only range of the spectrum for which 
Figure~\ref{fig:det} indicates we may be able to probe the $\alpha=2/3$ case.

\begin{figure*}
\centering
{\includegraphics[width=12.0cm]{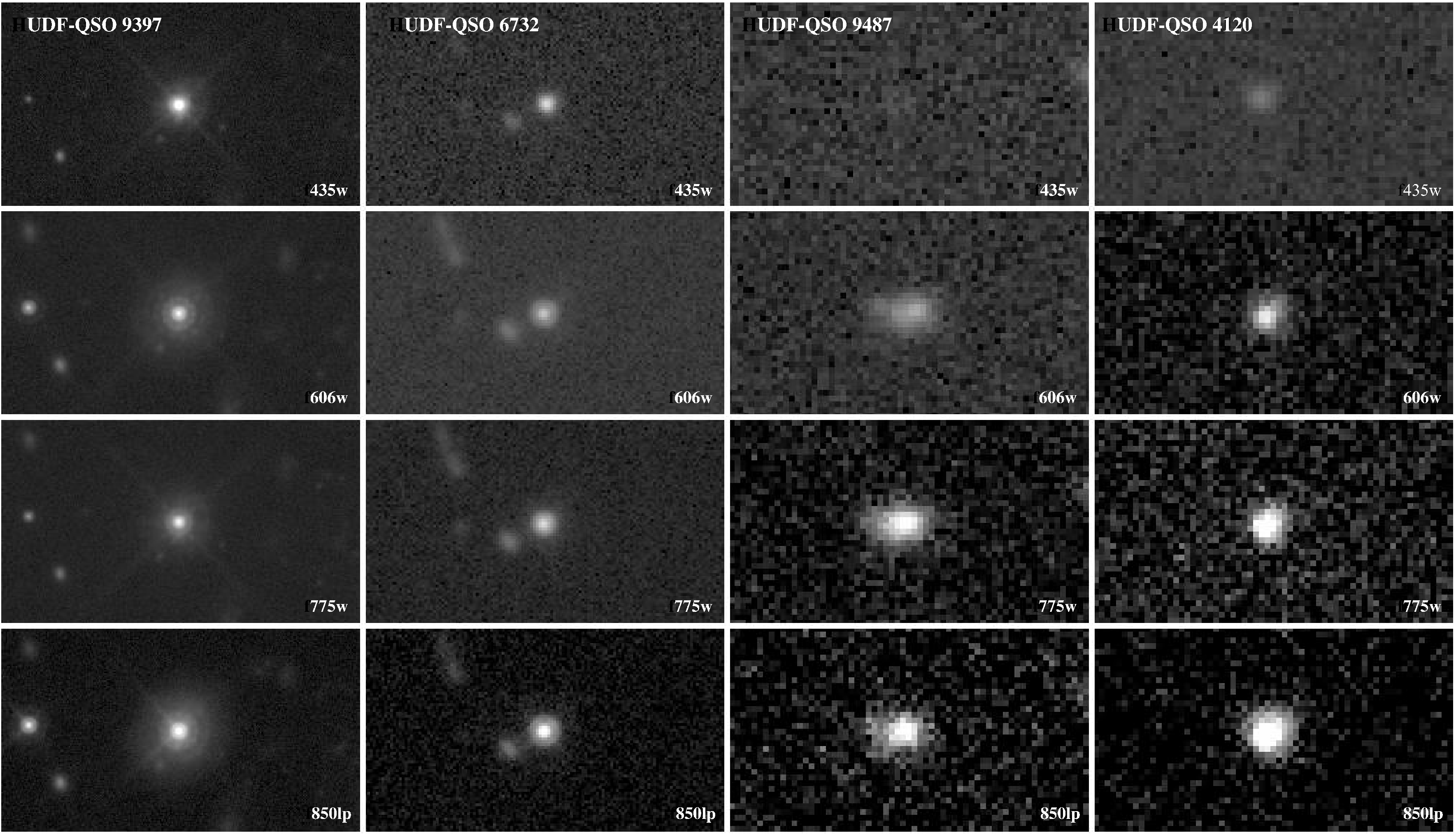}}
\caption{\label{fig:images} The four HUDF quasars studied in this paper, from left to right: 
9397 ($z=1.2$); 
6732 ($z=3.2$); 
9487 ($z=4.1$); 
4120 ($z=2.1$). 
Descending rows show the F435W (B), F606W (V), F775W (i) and F850LP (z) images.}
\end{figure*}

We turned next to the Hubble Ultra-Deep Field (UDF).
The UDF~\cite{HUDF} is the deepest image taken of the sky to date. The
$10^6$~s long exposure was taken with the ACS using a multi-angle dither
pattern to smooth out the PSF. A number of intermediate-to-high redshift
quasars have since been identified in this field, with redshifts provided by
the GRAPES project~\cite{pirzkal+04}. A list of quasars in the UDF was
published by~\cite{xu+07}. Four such quasars were chosen as suitable
candidates due to their relative isolation in the field (see
Table~\ref{tab:hst}). 
While we acknowledge the UDF data is not ideal for this task, since the 
sources will be inherently faint, and the data is assembled in a complicated 
stacking procedure which may introduce smoothing, it is only non-ideal in that 
it weakens our possibility of testing our null hypothesis: any smoothing 
effect serves to dilute the Strehl ratio further. 
%Of these, two were found to be
%suited to the present work, having sufficient signal to noise and no
%noticeable structure, and it is these objects that we study in depth here.
It is clear from the table that the advantage of using shorter wavelength 
B-band data more than compensates for the lower redshift of the sources 
when compared to the SDSS quasars examined by~\cite{ste07}.
Images of each UDF quasar in each band (B, V, i and z) are 
presented in Figure~\ref{fig:images}. 
It is noticeable from the original UDF images, that HUDF-QSO~6732 
appears somewhat blurred in the B filter (F435W). 
%However, one can see that
%blurring is present in the stellar images on the field also.
%For comparison, we present the TinyTim PSF's for each band in 
%Figure~\ref{fig:tt}, 
%the ideal azimuthally-symmetric PSF and stellar PSF's in 
%Figure~\ref{fig:stellar}.

\begin{figure*}
\centering
{\includegraphics[width=12.0cm]{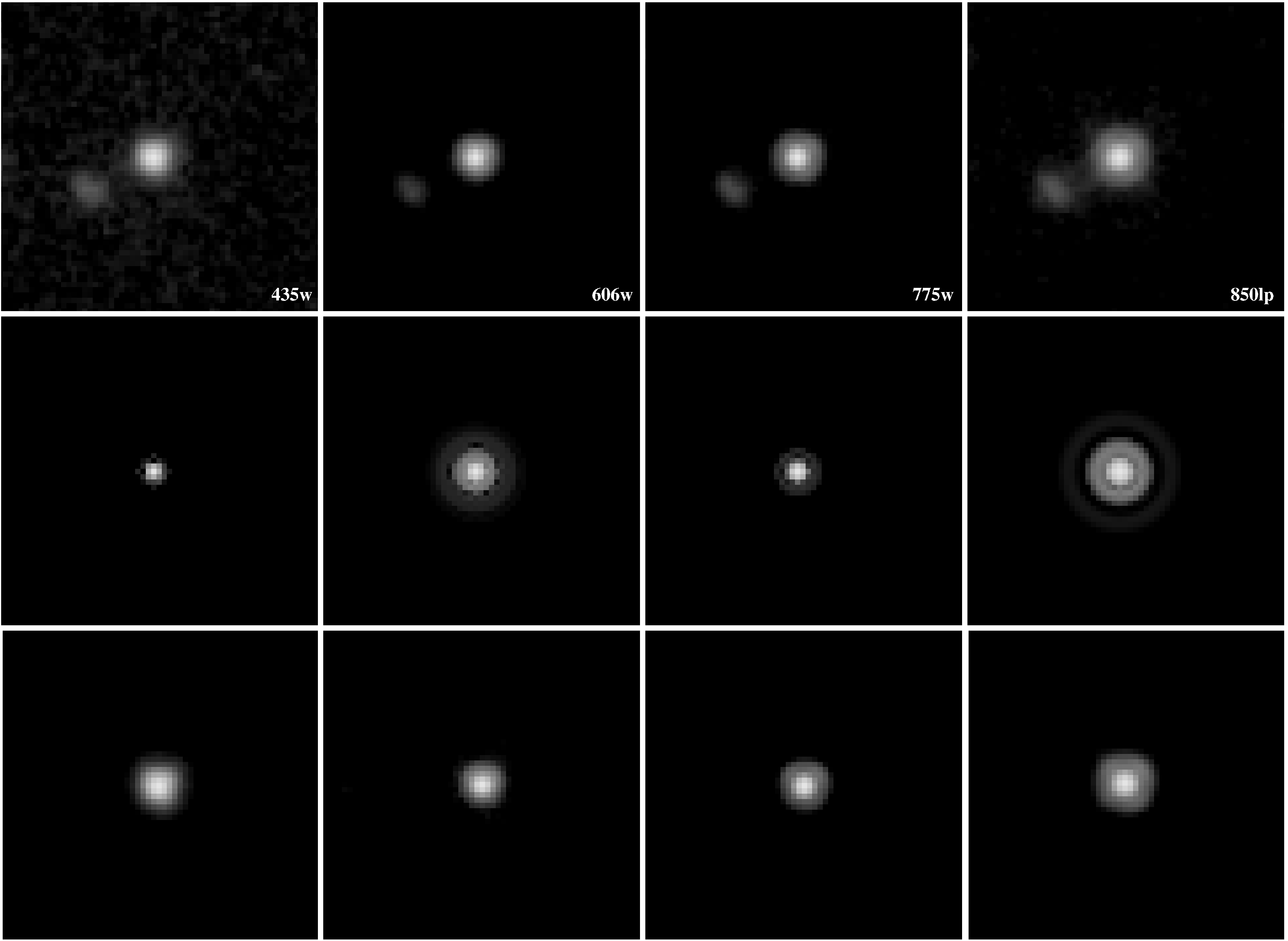}}
\caption{\label{fig:stellar} Top, from left-to-right: HUDF-QSO~6732 in 
F435W (B), F606W (V), F775W (i) and F850LP (z) bands. 
Middle: The ideal model PSF (1.2~m radius mirror with central hole of 0.35~m radius) for same filters. 
Bottom: Stellar PSF's for the same filters.}
\end{figure*}

As mentioned above, the
Strehl ratio is defined as the ratio of the image peak to that of the
diffraction spike of an unabberated telescope. The unaberrated PSF was
calculated assuming a primary mirror radius of 1.2~m with an inner radius of
0.35~m, convolved with the filter-instrument response function at a
resolution of 1\AA~to produce an unaberrated PSF for each filter.
These are shown along with stellar PSF's for each band in 
Figure~\ref{fig:stellar}.
However, as discussed by~\cite{ste07}, \HST has a complicated non
azimuthally symmetric structure to its PSF due to the support structure of
the telescope itself. This is partially mitigated by the UDF observing
strategy which smoothes the PSF and makes it rounder. Furthermore, the
processing that goes into the UDF removes the distortion present in raw
images. 
%TinyTim PSF's are shown for reference in Figure~\ref{fig:tt}.
Note that the SNR of the quasars is too low to be able to obtain reliable
photometry on the individual raw frames, which could be accurately compared
to Tinytim models, as was done by~\cite{ste07}. 
We found it futile to model the UDF PSF using TinyTim due
to the complicated stacking process (``drizzling''). Instead we resort to
stellar PSF stars to produce ``corrected'' Strehl ratios (after Steinbring
2007). While we cannot control the SED of the stars, we can perform repeat
experiments using multiple stars from across the field.

\begin{table}
\begin{center}
\caption{\label{tab:delI} Sample Uncertainties in $I$ }
\small
\begin{tabular}{lllr}
\hline
$z$ & $\delta z$ & $ \alpha$ & $\delta I(z, \alpha)/I(z,\alpha)$ \\  
\hline
\hline
1.00 & 0.01 & 0.50 & 0.0104 \\
1.00 & 0.01 & 0.60 & 0.0101 \\
1.00 & 0.01 & 0.67 & 0.0099 \\
5.00 & 0.01 & 0.50 & 0.0018 \\
5.00 & 0.01 & 0.60 & 0.0013 \\
5.00 & 0.01 & 0.67 & 0.0011 \\
5.00 & 0.05 & 0.50 & 0.0090 \\
5.00 & 0.05 & 0.60 & 0.0067 \\
5.00 & 0.05 & 0.67 & 0.0055 \\
%5.00 & 0.10 & 0.50 & 0.0180 \\
%5.00 & 0.10 & 0.60 & 0.0134 \\
%5.00 & 0.10 & 0.67 & 0.0109 \\ 
\hline
\end{tabular}
\end{center}
\end{table}

\begin{table*}
\begin{center}
\caption{\label{tab:hst} Quasar pointings with high $I(z,\alpha)/\lambda_o$ 
(see Eq.~\ref{eq3}) values in the \HST archive. 
$I(z,\alpha)/\lambda_o$ is a measure of the number of wavelengths 
travelled by the light detected at wavelength $\lambda_o$.
All calculations in this table assume $\alpha=2/3$.}
\scriptsize
\begin{tabular}{lrllrrr}
\hline
Name&$z$&Instrument&Filter&$\lambda_o$&$I(z,\alpha)$&$I(z,\alpha)/\lambda_o$\\
& & & &[$\mu$m]& &[$\mu$m$^{-1}$]~~~\\
\hline
\hline
SDSS~J1044--0125\cite{richards+04,ste07}&5.80&ACS/HRC&F850LP&0.91&6.98&7.71\\
SDSS~J-0836+0054\cite{richards+04,ste07}&5.82&ACS/HRC&F850LP&0.91&6.99&7.72\\
SDSS~J-1306+0356\cite{richards+04,ste07}&5.99&ACS/HRC&F850LP&0.91&7.06&7.81\\
SDSS~J-1030+0524\cite{richards+04,ste07}&6.28&ACS/HRC&F850LP&0.91&7.19&7.95\\
SDSS~J-0913+5919\cite{ste07,richards+06}&5.11&ACS/HRC&F775W&0.77&6.62&8.60\\
SDSS~J-2228-0757\cite{ste07,richards+06}&5.14&ACS/HRC&F775W&0.77&6.64&8.62\\
SDSS~J-1208+0010\cite{ste07,richards+06}&5.27&ACS/HRC&F775W&0.77&6.71&8.71\\
SDSS~J-0231-0728\cite{ste07,richards+06}&5.41&ACS/HRC&F775W&0.77&6.78&8.81\\
SDSS~J1148+5251\cite{white+05}&6.42&ACS/WFC&FR716N-7220&0.72&7.26&10.08\\%White
''''&''''&''''&FR914M-9050&0.90&''''&7.89\\%White
SDSS~J1148+5251\cite{stia03}&6.42&ACS/WFC&F775W&0.77&''''&9.42\\%Stiavelli -- no published???
''''&''''&''''&F850LP&0.91&''''&8.02\\
SDSS-J1048+4637\cite{stia03}&6.23&ACS/WFC&F775W&0.77&7.17&9.32\\
''''&''''&''''&F850LP&0.91&''''&7.93\\
SDSS-J1630+4012\cite{stia03}&6.05&ACS/WFC&F775W&0.77&7.09&9.21\\
''''&''''&''''&F850LP&0.91&''''&7.84\\
4C~45.51\cite{kukula+01}&1.992&NICMOS&F165M&1.65&3.01&1.82\\
4C~45.51\footnote{Floyd et al. (in preparation)}&1.992&WFPC2&F814W &0.81&3.01&3.69\\
223338-603329\cite{savaglio98}&2.24&STIS/FUVMAMA&F25QTZ&0.16&3.56&22.27\\
HUDF-QSO 4120\cite{xu+07}&2.095&ACS/WFC&F435W&0.43&3.25&7.52\\%QSO4
''''&''''&''''&F606W&0.59&''''&5.50\\%QSO4
''''&''''&''''&F775W&0.77&''''&4.22\\%QSO4
''''&''''&''''&F850LP&0.91&''''&3.59\\%QSO4
HUDF-QSO 6732\cite{xu+07}&3.193&ACS/WFC&F435W&0.43&5.08&11.76\\%QSO2
''''&''''&''''&F606W&0.59&''''&8.60\\%QSO2
''''&''''&''''&F775W&0.77&''''&6.60\\%QSO2
''''&''''&''''&F850LP&0.91&''''&5.61\\%QSO2
HUDF-QSO 9397\cite{xu+07}&1.225&ACS/WFC&F435W&0.43&0.60&1.39\\%QSO1
''''&''''&''''&F606W&0.59&''''&1.02\\%QSO1
''''&''''&''''&F775W&0.77&''''&0.78\\%QSO1
''''&''''&''''&F850LP&0.91&''''&0.67\\%QSO1
HUDF-QSO 9487\cite{xu+07}&4.094&ACS/WFC&F435W&0.43&5.96&13.79\\%QSO3
''''&''''&''''&F606W&0.59&''''&10.1\\%QSO3
''''&''''&''''&F775W&0.77&''''&7.74\\%QSO3
''''&''''&''''&F850LP&0.91&''''&6.58\\%QSO3
\hline
\end{tabular}
%\footnotetext{(a)}{\cite{richards+04}} 
%\footnotetext{(b)}{\cite{ste07} }
%\footnotetext{(c)}{\cite{richards+06}} 
%\footnotetext{(d)}{\cite{white+05} }
%\footnotetext{(e)}{\cite{stia03} }
%\footnotetext{(f)}{\cite{kukula+01} }
%\footnotetext{(g)}{Floyd et al. (in preparation)} 
%\footnotetext{(h)}{\cite{savaglio98} }
%\footnotetext{(i)}{\cite{xu+07}}
\end{center}
\end{table*}

We determined the centroid of each image and coaligned quasar and stellar
PSF images to the centre of the nearest pixel. We measured the Strehl ratio
for each quasar and stellar PSF image, with respect to the ideal, azimuthally 
symmetric PSF -- see Table~\ref{tab:SR1}.
We then divide the quasar Strehl ratio by each 
Stellar PSF Strehl ratio to produce a
corrected Strehl, $S_{\mathrm QSO}/S_{\mathrm Star}$, after~\cite{ste07}.
This is found to be equivalent to measuring the Strehl ratio with respect
to the stellar PSF, $S_{\mathrm QSO-Star}$ -- see table~\ref{tab:SR} --
which is as expected for well-centered images.
Referring to Eq.~\ref{eq5}, we get
\begin{equation}
S_\mathrm{Star} = \exp [-\sigma_I^2],
\label{eq10}
\end{equation}
so
\begin{equation}
S_\mathrm{QSO}/S_\mathrm{Star} = S_M \exp [-\sigma_{\phi}^2] = S_M S_{\phi}.
\label{eq11}
\end{equation}

\begin{figure*}
\centering
{\includegraphics[width=6.5cm]{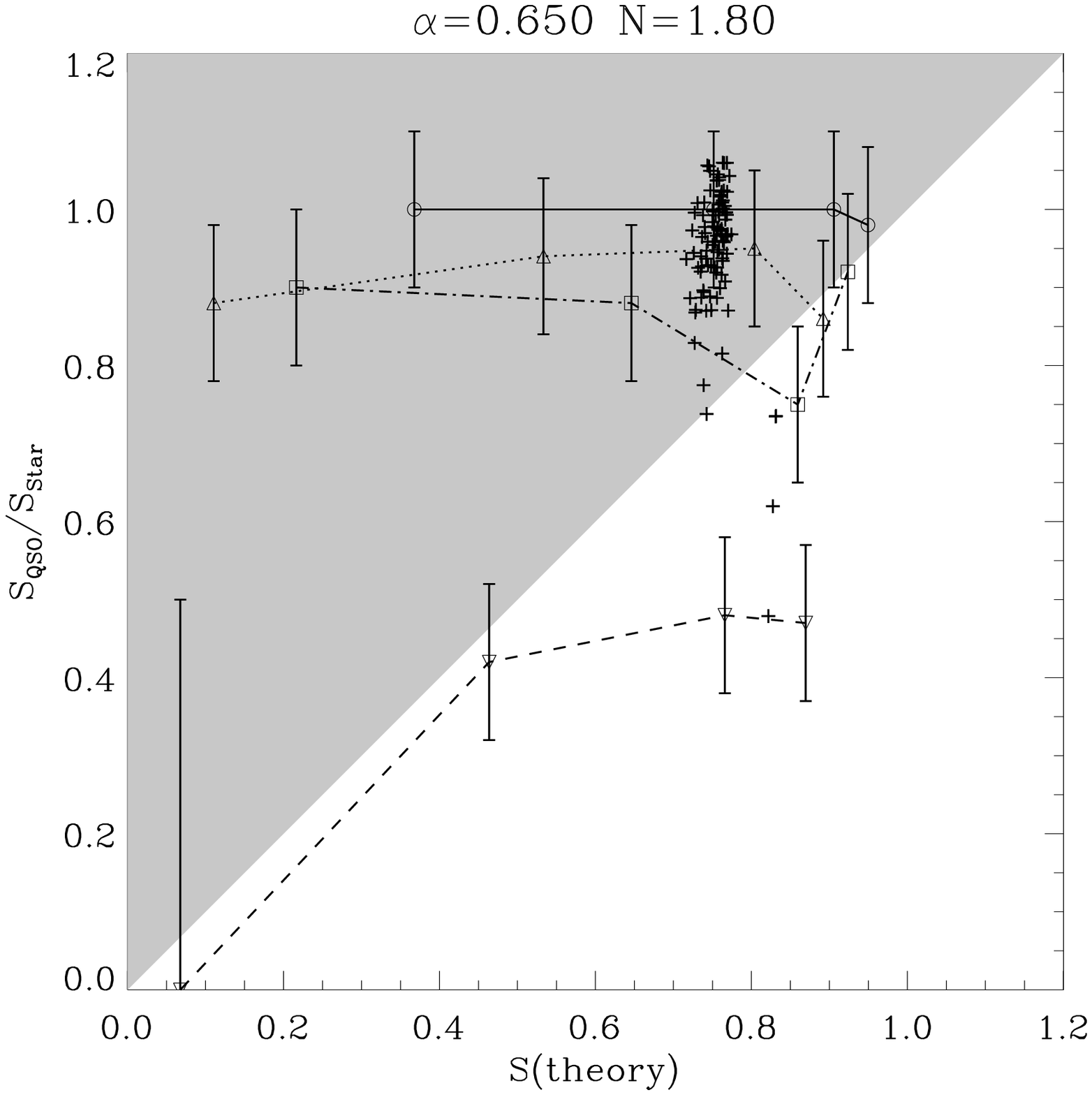}}
{\includegraphics[width=6.5cm]{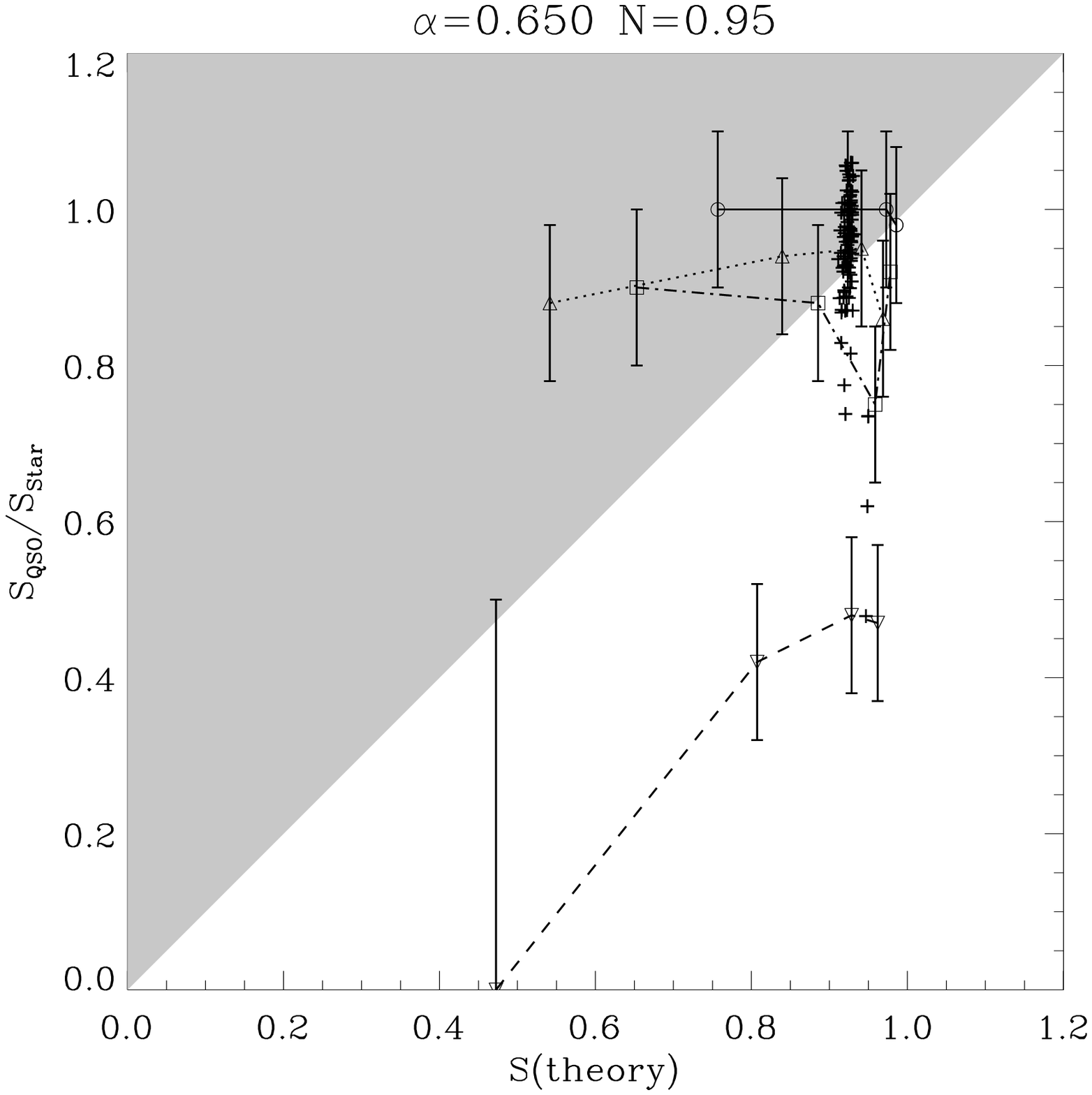}}

\vspace{2mm}
{\includegraphics[width=6.5cm]{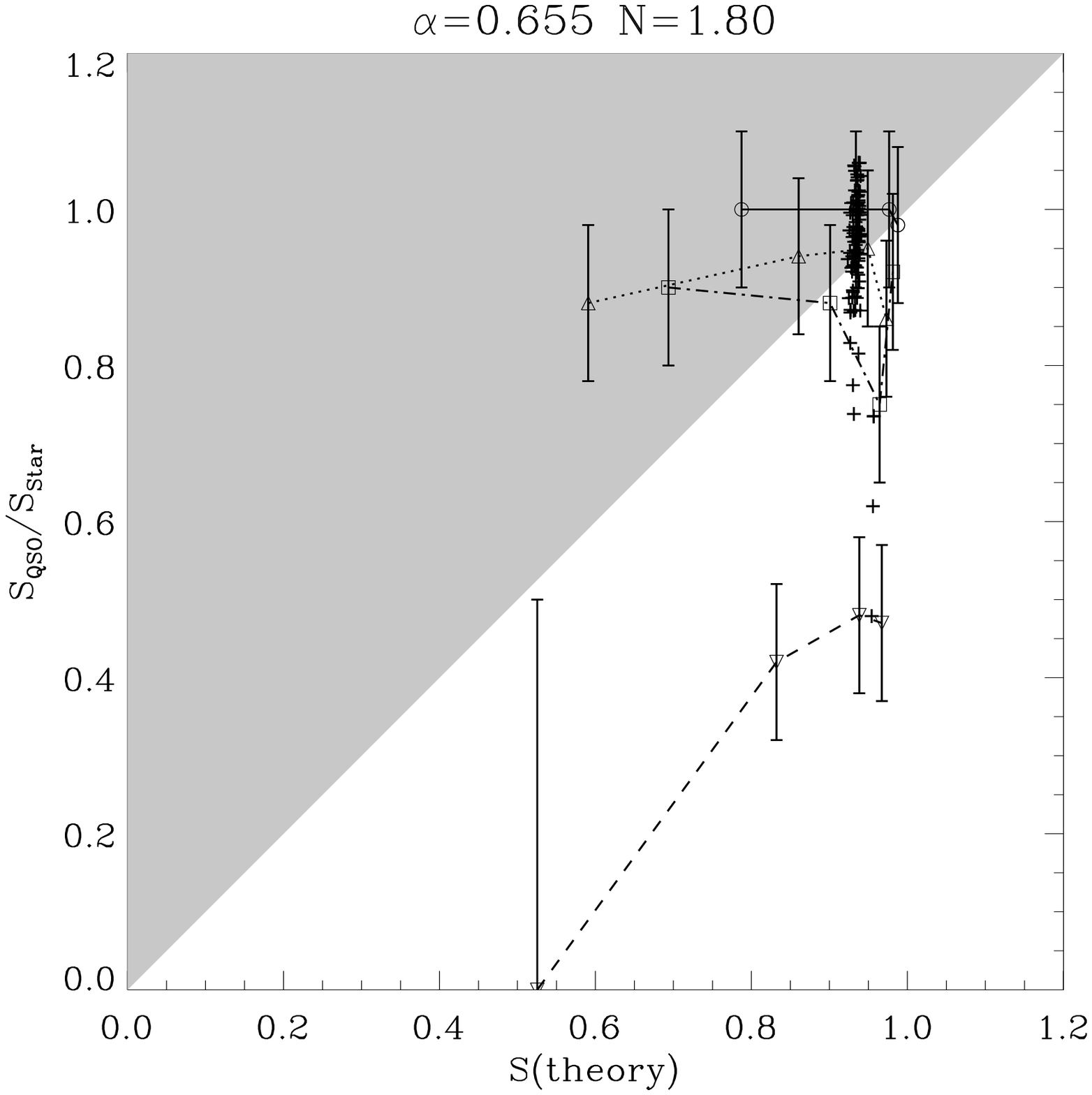}}
{\includegraphics[width=6.5cm]{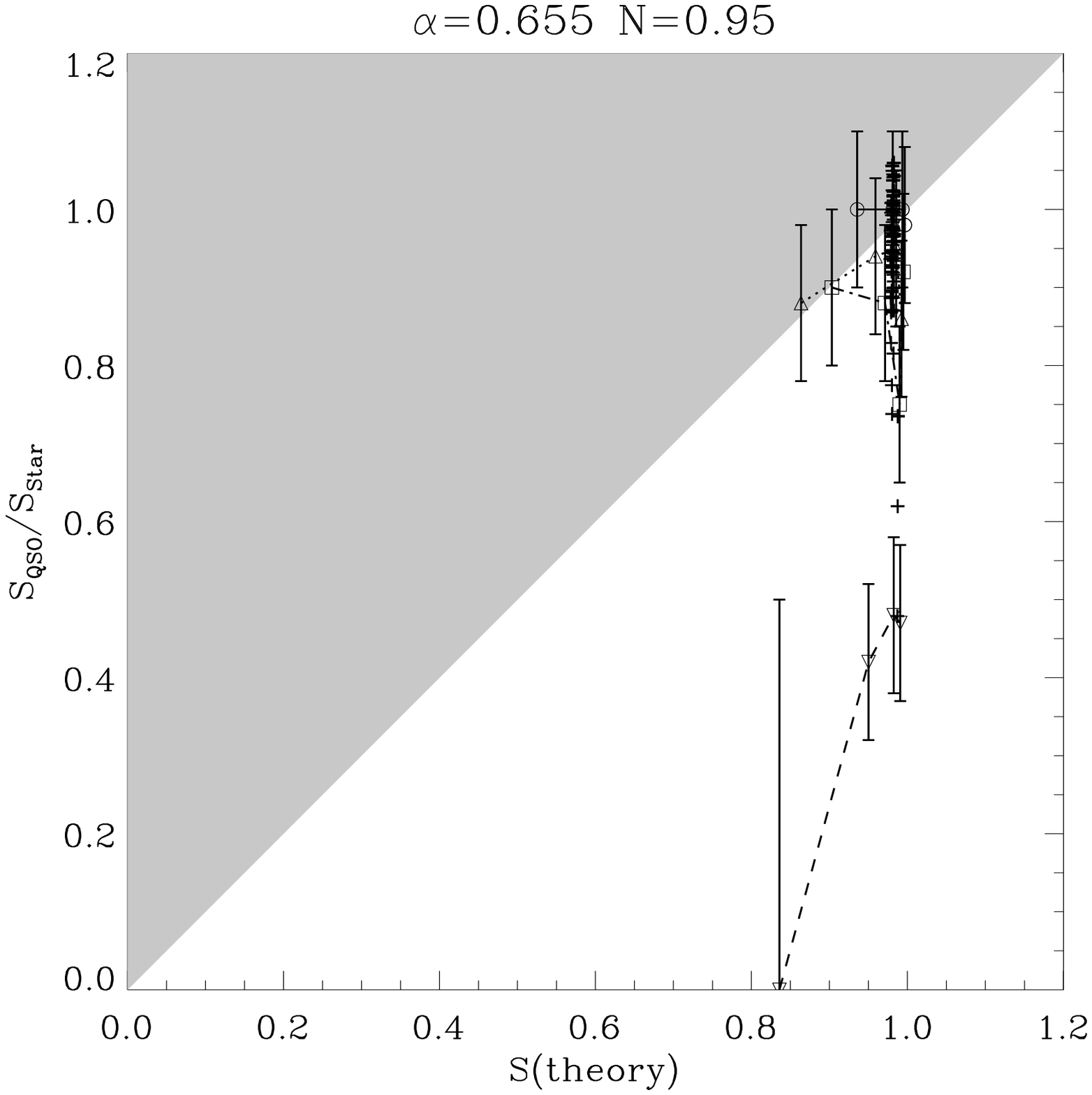}}

\vspace{2mm}
{\includegraphics[width=6.5cm]{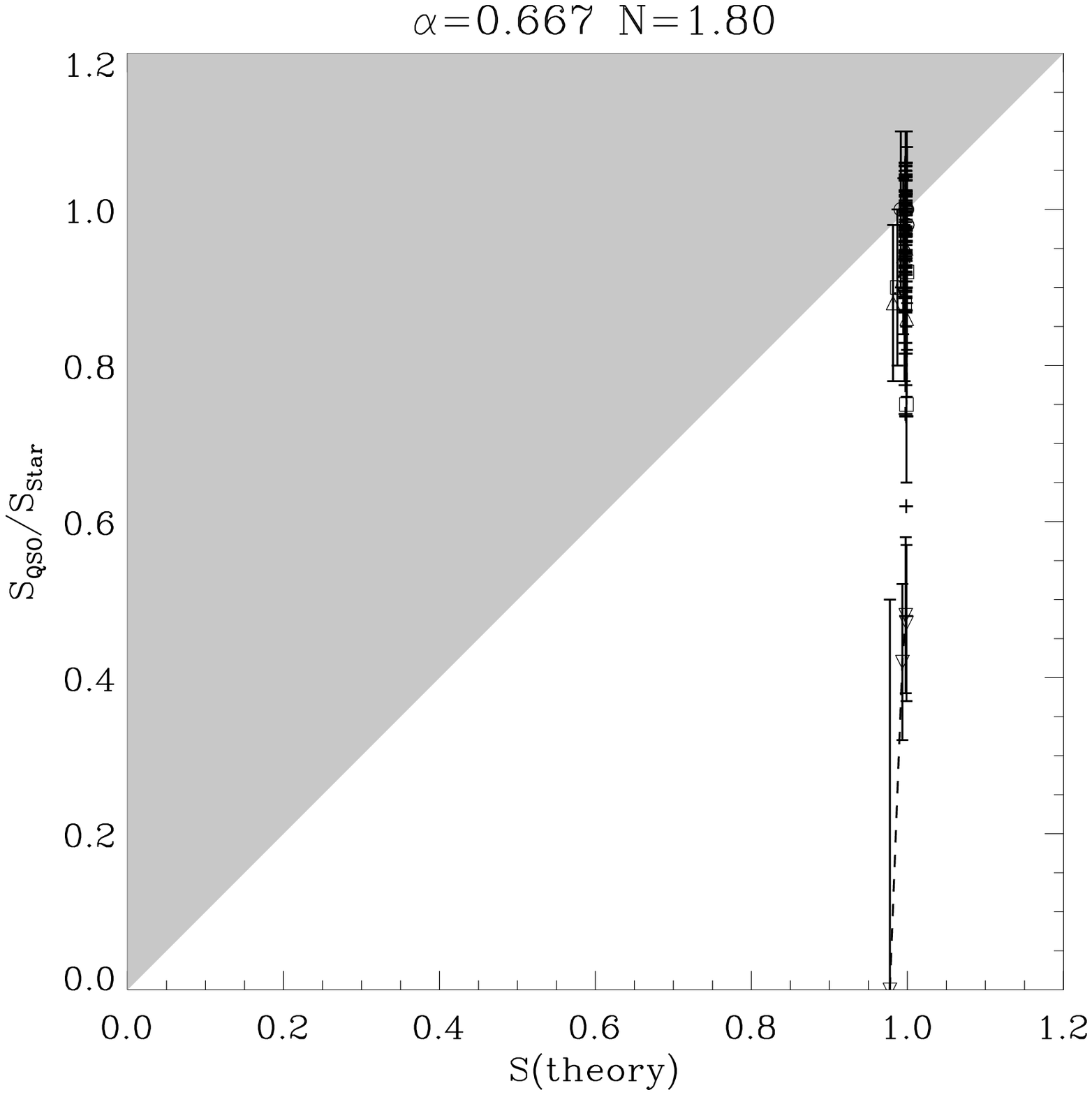}}
{\includegraphics[width=6.5cm]{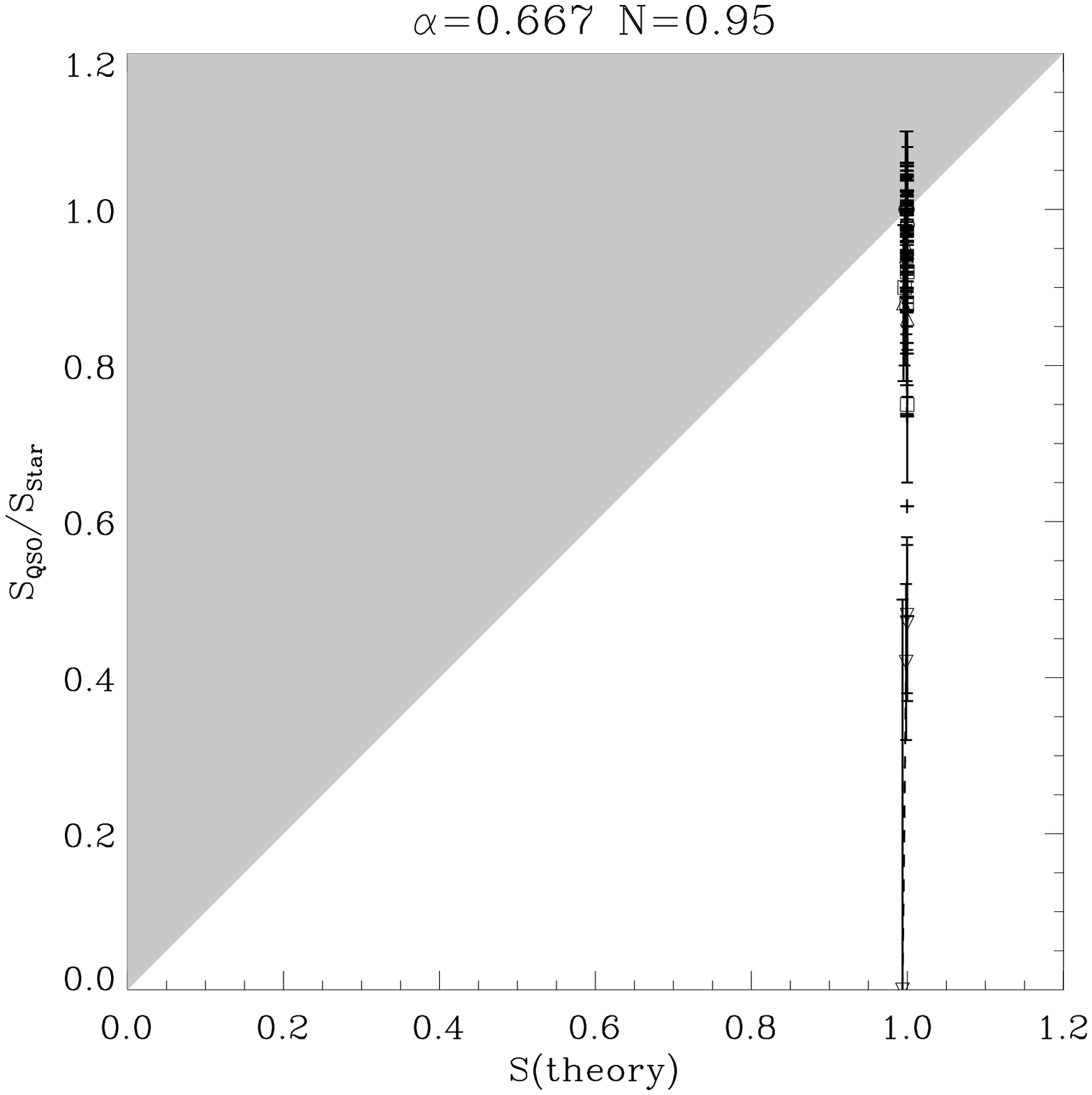}}
\caption{\label{fig:res} Measured Strehl ratio $S$ for the four UDF quasars 
(9397: circles \& solid line, 6732: up-pointing triangles \& dotted line, 
9487: squares \& dot-dashed line, 4120: down-pointing triangles \& dashed line)
against theoretical Strehl ratio $S_\phi$ (Eq.~\ref{eq:SR}), for various spacetime 
foam models. 
The shaded area indicates $S>S_\phi$: Such observations exclude a given 
spacetime foam model, $(\alpha, N)$.
Crosses indicate data points from~\cite{ste07}.
We exclude all models with $\alpha<0.65$ since we clearly see quasars that 
are less blurred than is predicted by such models (top).
We do in fact see a small amount of blurring in our UDF quasars, consistent 
with the magnitude of blurring expected for $\alpha=0.655$ (middle). 
No degradation should be seen for $\alpha=2/3$ (bottom) for the wavelengths 
and distances probed so far by the \HST archive ($\lambda > 400$~nm).}
\end{figure*}

We find that the stars produce quite stable Strehl ratios 
in each filter, across the field (varying at the $\ltsim5$\% level), while 
the quasars are found to vary in Strehl ratio considerably in a given filter. 
We measured all Strehl ratios using a range of photometric radii, from just 
$\theta_{\mathrm Airy}$, up to 0\farcs5, and found results to be 
consistent within the errors at each photometric radius. 
We adopt 0\farcs25 as our photometric radius.
Errors are dominated by systematics, estimated conservatively at 
$\sim \pm 10$\% as may be inferred from Figure~\ref{fig:res} where a few points have
$S_\mathrm{QSO}/S_\mathrm{Star}$ ratios $\sim 1.1$. 
%from the one case where we get a 
%Strehl ratio significantly greater than unity.
Repeat measurements at differing photometric radii gave far lower variation 
$\sim1$\%.

The Strehl ratios of each quasar from the UDF with respect to the UDF stellar 
PSF's are plotted in Figure~\ref{fig:res}, against the theoretical Strehl ratio 
obtained for various values of $\alpha$ and $N$. Any measured degradation 
above the 1:1 line (shaded region) rules out that spacetime foam model, 
since the quasars are not sufficiently blurry. That is, the image is degraded 
{\em less} than predicted by the theoretical model.
Points below the 1:1 line may indicate structure 
in the sources, i.e., the blurriness is above what we would expect for a given 
spacetime foam model and, thus, does not test the theory. 

Some blurring is seen, consistent with the level expected if $\alpha=0.655$. 
All models with $\alpha\leq 0.65$ are excluded by the UDF observations.

It is straightforward to disprove a given model, by finding an unblurred quasar 
just a little more distant or at shorter wavelengths (higher in the 
plot). Points lower in the plot may imply structure in the source. However, with 
the data we currently have, the alternate explanation, namely the combination of
limited Strehl ratio and the signal to noise of the UDF quasar observations, is 
equally plausible. We are clearly not yet probing the $\alpha=2/3$ scenario.
Larger statistical samples eventually may allow us to disentangle 
blurring due to structure from that due to the effects of spacetime foam.

\begin{table} \begin{center} \caption{\label{tab:SR1} Strehl ratios for  UDF
quasars (with identifier from~\cite{xu+07}) and stars  (with UDF coordinate
positions) in each filter,  with respect to the idealized, azimuthally symmetric
PSF. Unmeasurable values occur where the image is saturated.} \scriptsize
\begin{tabular}{lrrrr} \hline Source & F435W & F606W & F775W & F850LP \\ \hline
\hline QSO 9397 & 0.31 & 0.40 & 0.60 & 0.60 \\  QSO 6732 & 0.24 & 0.37 & 0.56 &
0.52 \\  QSO 9487 & ... & 0.17 & 0.28 & 0.29 \\  QSO 4120 & 0.25 & 0.35 & 0.44 &
0.56 \\  Star A (3026,3408) & 0.27 &	... & ... & 0.64\\ Star B (5692,4830) &
0.28 &	... & ... & 0.63\\ Star C (5006,2496) & 0.28 & 0.40 & 0.59 & 0.62\\ Star
D (8459,6278) & 0.28 & 0.34 & ... & 0.62\\ Star E (1231,5437) & 0.29 & 0.36 &
0.58 & 0.63\\ \hline \end{tabular} \end{center} \end{table}

\begin{table}
\begin{center}
\caption{\label{tab:SR}
Strehl ratios for the UDF quasars (identifiers from~\cite{xu+07}) 
with respect to stellar PSF's, $S_\mathrm{QSO-Star}$, using a photometric aperture of 0\farcs25.}
\scriptsize
\begin{tabular}{lrrrr}
\hline
Source & F435W & F606W & F775W & F850LP \\
\hline
\hline
9397 & $1.10\pm0.1$ & $1.02\pm0.1$ & $1.02\pm0.1$ & $0.98\pm0.1$ \\
6732 & $0.88\pm0.1$ & $0.94\pm0.1$ & $0.95\pm0.1$ & $0.86\pm0.1$ \\
9487 & $0.00\pm0.5$ & $0.42\pm0.1$ & $0.48\pm0.1$ & $0.47\pm0.1$ \\
4120 & $0.90\pm0.1$ & $0.88\pm0.1$ & $0.75\pm0.1$ & $0.92\pm0.1$ \\
\hline
\end{tabular}
\end{center}
\end{table}

%%%%%%%%%%%%%%%%%%%%%%%%%%%%%%%

%VI. {\bf Discussion and Conclusion}
\section{Discussion and Conclusion}
\label{sec-conc}
Using the UDF data,
we see evidence for an increased blurring (decreased optically corrected
Strehl ratio) with increasing $I(z,\alpha)/\lambda_o$. This is likely due to 
masking or source structure in the majority of cases, but we note that the blurring 
is at a level consistent with a spacetime foam model with $\alpha=0.655$. All 
models with $\alpha<0.65$ are excluded by the UDF observations.

We are not yet probing the $\alpha=2/3$ regime, however; the UDF data is 
non-optimal for this purpose as they are not at a wavelength short enough 
to exclude the holographic model. 
Indeed, we see obvious blurring in the quasar images, in some cases due either to 
structure or masking. Nevertheless, the UDF observations provide the strongest 
constraints that currently exist on spacetime foam.

We have shown that the current \HST archive does not contain observations that 
can more tightly constrain spacetime foam models. To do this, we would need 
observations in the ultraviolet. While one such observation currently 
exists, it was made by the STIS instrument, which has a triangular PSF with no 
Airy ring, making the analysis we detail in section~\ref{sec-HST} impossible. 
The measurement can be done either by the new WFC3 instrument, which 
can access the very edge of the $\alpha=2/3$-sensitive region of the parameter 
space at the blue end of its spectral range (using the F225W, F275W 
or F300X filters), or alternately by the ACS SBC, which can probe to 100 nm. 
While both instruments under-sample the PSF in the ultraviolet, each has its 
advantages. The WFC3 is considerably more sensitive, so that higher dynamic 
ranges can be probed in a shorter exposure time. However, at the short 
wavelengths probed by the SBC the anticipated blurring is so 
strong that the under-sampling may not matter.

The two main difficulties in probing high redshift quasars at such short 
wavelengths are scattering by their extended haloes, and absorption by 
intervening HI and HeII. The former is the most likely explanation for 
any observed structure in the quasars, should it be observed, but can be 
distinguished from the effects of spacetime foam using polarimetric measurements 
(e.g., \cite{Vernet01,Young09}).  Furthermore, we do not 
expect scattering to dominate the flux for type 1 quasars, even at 
wavelengths $\sim1215$\AA, since \cite{Vernet01} found a wavelength 
independent scattering law for obscured quasars.
The latter is the more significant problem, as we are already photon 
starved at restframe wavelengths shortward of Ly alpha, and at redshifts 
$\gtrsim 4.3$ both HI and HeII absorption will affect the opacity at 
wavelengths $<1215$\AA~ in the quasar frame \cite{Meiksin06, Madau95}.
Nevertheless, in the absence of diffraction-limited X-ray imaging in the 
foreseeable future, rest-frame far-UV imaging represents the best hope 
for probing the $\alpha=2/3$ regime, although we admit we would be lucky to 
find a quasar sightline that unambiguously disproved the $\alpha=2/3$ case 
by detecting no blurring or structure. In the rather more likely event of 
structure being detected, it would take a series of deep and expensive 
observations to establish that the blurring or structure was not due to 
scattering.
The best sources to investigate are the high resdshift quasars from the 
Richards (SDSS) catalogue that are known to be isolated and to show no 
structure in their z or I band images. All the UDF quasars show 
companions or similar nearby structure.

%While in It is difficult to attribute this to the host galaxy, as one would expect to
%detect any extended flux in the red preferentially. Indeed, it seems
%unlikely that the blurring is intrinsic to the quasar, as every expected
%selection effect works the other way.

%Notes: (??? FOR DISCUSSION/CONCLUSIONS???)

%1. For future tests with \HST, it would be advantageous to follow an
%observational program similar to that adopted for the study of quasar host
%galaxies (e.g.~\cite{floyd+04}), in which a PSF star is chosen for its
%similar SED to quasars at the redshift of interest, and is observed close in
%time to the quasar and on the same part of the imaging chip in question.
%Need to go into the NUV in order to really probe the parameter space for 
%$\alpha=2/3$. ACS/SBC is only option, although WFC3 at lowest wavelengths 
%would provide good baseline: Important to follow target all the 
%way from blue to UV

%???Might also try to look at specific BLR emission lines (CIV, Hbeta) using
%e.g. a tuneable filter.

%We clearly need more data! Revisiting the UDF quasars with a normal
%observing strategy might work.

%For discussion:

%VLTI Spectro-Imager: potential application?\\

As already noted at the end of section~\ref{sec-obstest}, one of the most exciting prospects is
that the fully operational VLTI will allow us to test the Holographic model.
In fact, as shown by Figure~\ref{fig:det}, due to the combination of its baseline size
plus wavelength, the VLTI emerges as the best overall current or 
planned instrument for the testing of spacetime foam models. 
The instrumental setup to use would
be either MIDI (at wavelengths longer than 5 $\mu$m) or PRIMA (at shorter
wavelengths). This would be a rather difficult observation, because
currently, VLT interferometry is limited to the very brightest sources: 
in $K$ band, the practical limit is $K=10$, although that can be extended 
down to $K=15$ if a bright star of $K=10$ or brighter is very close to 
(within a few arcseconds) the quasar (F. Delplancke, priv. comm.). 
The presence of such a bright point source allows one to 
'phase-reference' the observation so that the visibility of the
fringes can be tracked with high precision. This is a significant
difficulty, as the number of quasars brighter than $K=10$ is very 
small -- in fact, a search in SIMBAD revealed only one such object, 
namely 3C 273, which is at the modest
redshift of $z=0.158$. If one looks for bright stars within 40 arcsec, one
finds a few other quasars with bright stars nearby, specifically PKS
0435-300, HS 1227+4641, OK 568, AH 26 and 2GZ J011339-3343. All of 
these are at the lower end of the useful range as far as the quasar 
brightness is concerned, and also, two or three are too far north for VLTI 
observations (HS 1227+4641, OK 568, possibly AH 26).

The initial conclusion would then be that this is a difficult observation to
do and not many sources are available to do it. However, one can use 
targets of opportunity to explore further. For example, bright blazars in 
outburst can reach up to $K \sim 12$ and in extreme cases $K \sim 11$, 
making them possibly useful for these observations, particularly if the 
future holds an increase in sensitivity. 
Another possibility is to use the afterglows of bright gamma-ray bursts, 
which have now in at least one case gotten considerably brighter
than 10th magnitude in K. 
Examples here include GRB 990123, which reached 
$V\sim 9$ in its prompt phase (Galama et al. 1999),
and GRB 060607A (Ziaeepour et al. 2008), which was $B \sim 13$ 
in its prompt phase.
However, considering that GRBs decline precipitously in brightness within
minutes, their utility for VLTI observations is likely quite limited.

We conclude with a succinct summary.  In this paper we elaborate on
an earlier 
proposal (CNvD, 2006) to detect spacetime foam by looking for
seeing disks in the images of distant quasars and AGNs.  Assuming isotropic
fluctuations, we argue that spacetime in effect creates a seeing disk whose
angular diameter is $\sim \delta \psi$ and that the effect is expected to show
up in interferometric fringe pattern as decreasing fringe visibility and
reduction in Strehl ratios as soon as $\delta \psi \sim \lambda/ D$.  A model of
spacetime foam is disproved if images of a distant source
do not exhibit the
blurring predicted by the model.  Thus far, images of high-redshift
quasars
from HUDF provide the most stringent tests of spaceteime foam models. 
The random-walk model is convincingly ruled out while the holographic model is
still viable.  We anticipate that the latter model can be tested when the
VLTI is fully operational; it is also possible that observations with HST in the
ultraviolet may accomplish this goal.

\acknowledgments
YJN is supported in part by the US Department of Energy under contract
DE-FG02-06ER41418.
DJEF acknowledges the support of a Magellan Fellowship from Astronomy
Australia Limited, and administered by the Anglo-Australian Observatory.
We thank F. Delplancke for useful email correspondence.
We also thank R. Fosbury for useful comments.  YJN thanks 
A. Glindemann for the hospitality extended to him while he 
was visiting the Headquarters of ESO.
This paper is partly based on observations with the NASA/ESA 
{\it Hubble Space Telescope},
obtained at the Space Telescope Science Institute.
%We thank the referee whose comments and questions led to helpful clarifications of this work

%\bibliographystyle{astron}

%\bibliography{full_lib}
%\bibliography{biblio}
%

\end{document}